\documentclass[%
 reprint,
superscriptaddress,
amsmath,amssymb,
 aps,
longbibliography]{revtex4-2}

\usepackage{graphicx}
\usepackage{dcolumn}
\usepackage{bm}
\usepackage{chemformula}
\usepackage{amsmath,amsthm,amssymb}
\usepackage{gensymb}
\usepackage{xr}
\usepackage{hyperref}
\usepackage{soul}

\makeatletter
\newcommand*{\addFileDependency}[1]{
  \typeout{(#1)}
  \@addtofilelist{#1}
  \IfFileExists{#1}{}{\typeout{No file #1.}}
}
\makeatother

\usepackage[mathlines]{lineno}

\begin{document}

\title{Halide perovskite artificial solids as a new platform to simulate collective phenomena in doped Mott insulators}

\author{Alessandra Milloch}
\email[]{alessandra.milloch@unicatt.it}
\affiliation{Department of Mathematics and Physics, Università Cattolica del Sacro Cuore, Brescia I-25133, Italy}
\affiliation{ILAMP (Interdisciplinary Laboratories for Advanced
Materials Physics), Università Cattolica del Sacro Cuore, Brescia I-25133, Italy}
\affiliation{Department of Physics and Astronomy, KU Leuven, B-3001 Leuven, Belgium}

\author{Umberto Filippi}
\affiliation{Italian Institute of Technology (IIT), Genova 16163, Italy}

\author{Paolo Franceschini}
\affiliation{CNR-INO (National Institute of Optics), via Branze 45, 25123 Brescia, Italy}

\author{Michele Galvani}
\affiliation{Department of Mathematics and Physics, Università Cattolica del Sacro Cuore, Brescia I-25133, Italy}

\author{Selene Mor}
\affiliation{Department of Mathematics and Physics, Università Cattolica del Sacro Cuore, Brescia I-25133, Italy}
\affiliation{ILAMP (Interdisciplinary Laboratories for Advanced
Materials Physics), Università Cattolica del Sacro Cuore, Brescia I-25133, Italy}

\author{Stefania Pagliara}
\affiliation{Department of Mathematics and Physics, Università Cattolica del Sacro Cuore, Brescia I-25133, Italy}
\affiliation{ILAMP (Interdisciplinary Laboratories for Advanced
Materials Physics), Università Cattolica del Sacro Cuore, Brescia I-25133, Italy}

\author{Gabriele Ferrini}
\affiliation{Department of Mathematics and Physics, Università Cattolica del Sacro Cuore, Brescia I-25133, Italy}
\affiliation{ILAMP (Interdisciplinary Laboratories for Advanced
Materials Physics), Università Cattolica del Sacro Cuore, Brescia I-25133, Italy}

\author{Francesco Banfi}
\affiliation{FemtoNanoOptics group, Université de Lyon, CNRS, Université Claude Bernard Lyon 1, Institut Lumière Matière, F-69622 Villeurbanne, France}

\author{Massimo Capone}
\affiliation{International School for Advanced Studies (SISSA), Trieste 34136, Italy}

\author{Dmitry Baranov}
\affiliation{Italian Institute of Technology (IIT), Genova 16163, Italy}
\affiliation{Division of Chemical Physics, Department of Chemistry, Lund University, P.O. Box 124, SE-221 00 Lund, Sweden}
 
\author{Liberato Manna}
\affiliation{Italian Institute of Technology (IIT), Genova 16163, Italy}

\author{Claudio Giannetti}
\email[]{claudio.giannetti@unicatt.it}
\affiliation{Department of Mathematics and Physics, Università Cattolica del Sacro Cuore, Brescia I-25133, Italy}
\affiliation{ILAMP (Interdisciplinary Laboratories for Advanced
Materials Physics), Università Cattolica del Sacro Cuore, Brescia I-25133, Italy}
\affiliation{CNR-INO (National Institute of Optics), via Branze 45, 25123 Brescia, Italy}

\begin{abstract}
The development of Quantum Simulators, artificial platforms where the predictions of many-body theories of correlated quantum materials can be tested in a controllable and tunable way, is one of the main challenges of condensed matter physics. Here we introduce artificial lattices made of lead halide perovskite nanocubes as a new platform to simulate and investigate the physics of correlated quantum materials. The ultrafast optical injection of quantum confined excitons plays a similar role to doping in real materials. We show that, at large photo-doping, the exciton gas undergoes an excitonic Mott transition, which can be mapped on the insulator-to-metal transition of the Hubbard model in a magnetic field. At lower photo-doping, the long-range interactions drive the formation of a collective superradiant state, in which the phases of the excitons generated in each single perovskite nanocube are coherently locked. Our results demonstrate that time-resolved experiments span a parameter region of the Hubbard model in which
long-range and phase-coherent orders emerge out of a doped Mott insulating phase. This physics is relevant for a broad class of phenomena, such as superconductivity and charge-density waves in correlated materials whose properties are captured by doped Hubbard models.
\end{abstract}

\maketitle

 \begin{figure}[t]
\includegraphics[width=8.5cm]{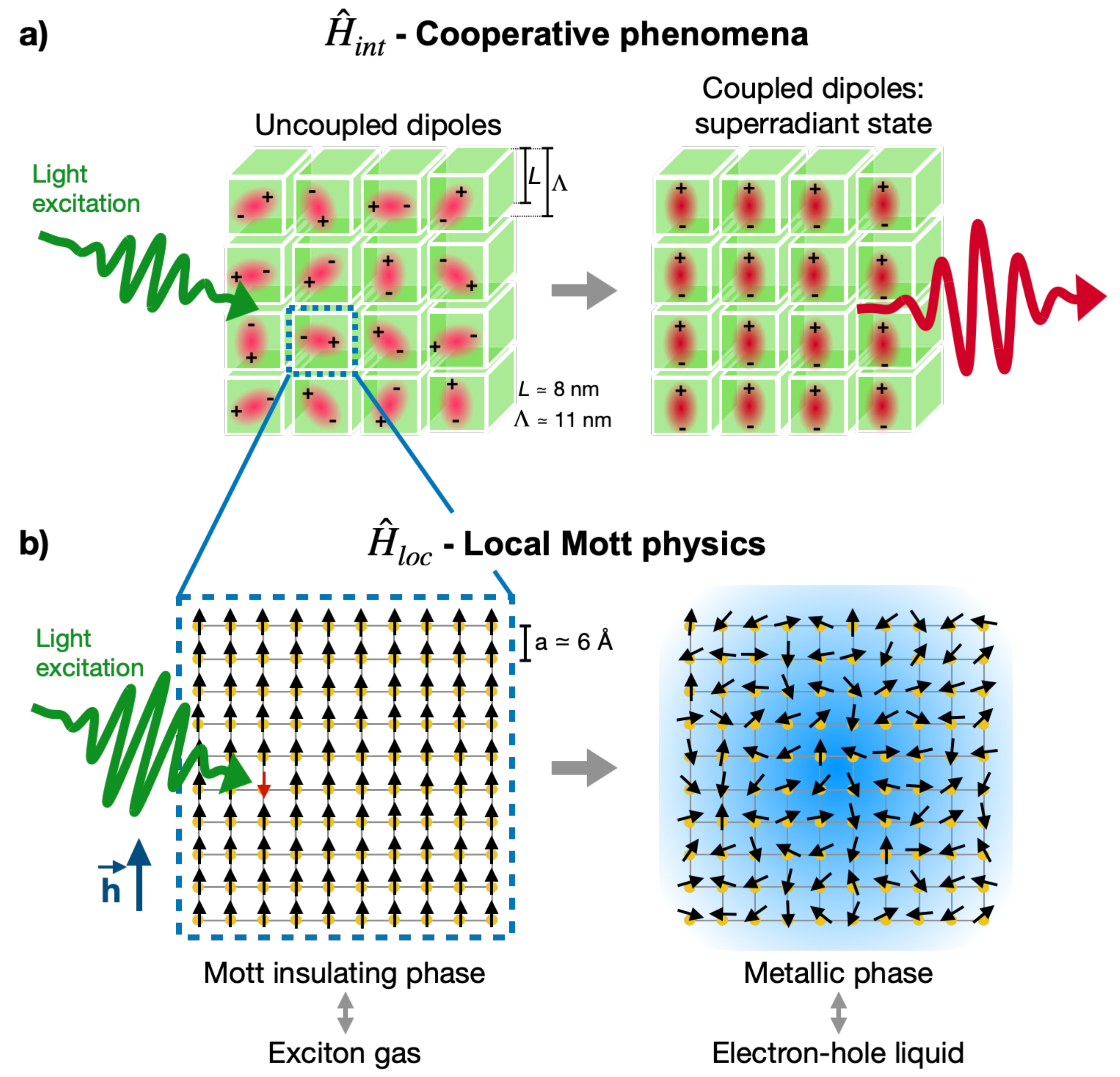}
\caption{Cartoon describing how perovskite quantum dot artificial solids  can be harnessed to simulate cooperative phases in the vicinity of the Mott insulator-to-metal transition. a) Cooperative effects - such as superradiance - emerge from long-range interactions $\hat{H}_{int}$ when coupled via an external light field. b) Local interactions $\hat{H}_{loc}$ govern the photo-induced transition from an exciton gas and an electron-hole liquid in a semiconductor, realizing a Mott transition that can be fully mapped onto a repulsive Hubbard model in a magnetic field $h$.}
\label{fig1: cartoon}
\end{figure}

\section{INTRODUCTION}
The paradigm of quantum simulations \cite{altman2021quantum,Cirac2012} has been pioneered by the development of ultracold-atom systems \cite{bloch2008many, bloch2012quantum, chien2015quantum, freeney2022electronic} and extended to solid state via nano- and hetero-structured \cite{altman2021quantum,Bernevig2006,Buluta_2009,Lagoin2022} devices and, more recently, twisted bidimensional materials \cite{cao2018correlated, cao2018unconventional, carr2020electronic,kennes2021moire}. 
An additional promising path consists in coupling a quantum material with the photons of a cavity \cite{Walther_2006,Haroche_1989}, which opens the possibility to optically drive and control the emergence of collective phenomena and long-range coherence. Intense efforts are currently dedicated to the development of photonics-based platforms aimed at replicating the many-body physics of quantum correlated materials. External optical control of the microscopic parameters entering the relevant Hamiltonian, such as doping, hopping, and interaction strength, is key to tackle open problems relevant for solid state physics. An important example is given by the Hubbard model, which is believed to capture the most fundamental properties of correlated materials, such as copper oxides \cite{Imada1998}. For large values of the on-site Coulomb repulsion $U$, the Hubbard model reproduces the correlation-driven metal-to-insulator Mott transition, thus capturing the insulating nature of copper oxide parent compounds. When a small number of free carriers is injected (doping) the model hosts the emergence of a wealth of long-range phases, such as charge density waves and high-temperature superconductivity, characterized by the macroscopic phase locking of fundamental incoherent fluctuations. The understanding of these phenomena in real materials is still a major challenge that is driving the search for artificial systems where to investigate in controlled ways the manifestation of collective phenomena in doped Mott insulators. 

In this work we introduce macroscopic lattices constituted by lead halide perovskite nanocubes as a new photonic platform to artificially implement the Hubbard model, which describes the local excitonic physics within each nanocube. At the same time, the long-range interactions among excitons in different nanocubes drive the emergence of collective phase-coherent states. We perform broadband time-resolved optical measurements and demonstrate the possibility of spanning different quantum phases - namely, the excitonic Mott insulating phase, the superradiant collective state, and the metallic electron-hole liquid phase - on the same artificial solid system by continuously tuning the light excitation intensity. The possibility to simulate and control the emergence of long-range ordered phases in lightly doped Mott insulators is key to tackle relevant problems in condensed matter physics, such as the onset of superconductivity, charge and spin order in correlated oxides. This new platform offers a new, low-cost and high-temperature alternative to state-of-the-art quantum simulators based on cold-atoms trapped into optical lattices \cite{Jordens2008,Greif2016,Kaiser2016,Asenjo2017,Ferioli2021,Ferioli2021b,Sierra2022,Masson2022}. 

\section{short and long-range interactions in halide perovskite artificial lattices}
\label{sec: hamiltonians}
Lead halide perovskites (chemical structure \ch{APbX_3}) are attracting considerable attention thanks to their many appealing electronic properties - such as strong excitonic resonances, tunable band gap, strong absorption and emission features in the visible range, high photoluminescence quantum yield - that make them promising materials for optoelectronic applications \cite{fu2019metal,green2014emergence,he2023rise}. Tunability of the optical properties can be achieved not only by halide exchange (\ch{X} = \ch{Cl}, \ch{Br}, and \ch{I}), but also by quantum confinement that is possible through the synthesis of nanometric cubes (NCs) with edge size $L$ comparable to the exciton Bohr radius \cite{shamsi2019metal,dey2021state,protesescu2015nanocrystals}. These quantum dots can self-organize into highly ordered three-dimensional superlattices, thus creating large-scale artificial solids - with the nanocube acting as the fundamental unit cell - whose properties, such as superlattice parameter $\Lambda$, hopping, exciton energies and symmetry of the unit cell, can be tuned by chemical means \cite{brennan2020superlattices,
cherniukh2021perovskite}.
In these systems, each individual nanocube can host a controllable number of quantum confined excitons \cite{protesescu2015nanocrystals,sercel2019quasicubic}, whose properties are determined by the local electronic interactions. Furthermore, the electromagnetic field can drive long-range interactions among excitons in different nanocubes, thus realizing the total Hamiltonian:
\begin{equation}
\label{eq: Hubbard}
    \hat{H} = \hat{H}_{loc}+\hat{H}_{int}
\end{equation}
where $\hat{H}_{loc}$ describes the local interactions within each single nanocube and is defined on the perovskite cubic lattice with periodicity $a$, whereas $\hat{H}_{int}$ describes the inter-unit cell interactions and is defined on the superlattice with periodicity $\Lambda > a$.

The $\hat{H}_{int}$ term (see Materials and Methods) is responsible for a rich family of collective phenomena, broadly indicated as superradiant \cite{Cong_2016}. 
As sketched in the cartoon in figure \ref{fig1: cartoon}a, superradiance occurs when the electromagnetic-field-driven interaction leads to the phase-coherence of $N$ quantum emitters. A typical manifestation is the collective emission of radiation (superfluorescence), which is both enhanced and faster than the emission from individual nanocubes, with the radiative rate scaling as $N^2$ for large $N$ \cite{Cong_2016,dicke1954coherence}.
Groundbreaking photoluminescence (PL) experiments recently reported evidence of superfluorescence effects in halide perovskites \cite{raino2018superfluorescence,findik2021high,krieg2020monodisperse,cherniukh2021perovskite,zhou2020cooperative,biliroglu2022room}. The main manifestations of this collective phenomenon are: i) the superlinear dependence of the emission amplitude with respect to the intensity of the exciting external field \cite{raino2018superfluorescence,findik2021high,krieg2020monodisperse,cherniukh2021perovskite}; ii) the emergence of a narrow red-shifted peak in the PL spectrum, which is assigned to the cooperative emission from a sub-population of nanocubes within a single superlattice \cite{raino2018superfluorescence,findik2021high,krieg2020monodisperse,cherniukh2021perovskite}. These cooperative effects are suppressed at high temperature in nanocube superlattices due to thermal noise that undermines quantum coherence \cite{mattiotti2020thermal}.

The local interactions described by $\hat{H}_{loc}$ can give rise to a transition from an exciton gas (EG) to a liquid of weakly interacting electrons and holes (EHL), achieved when a very large number of excitons is photo-injected in bulk and low-dimensional semiconductors \cite{palmieri2020mahan, chernikov2015population, bataller2019dense,schlaus2019lasing,yu2019room}.
This transition is believed to almost perfectly realize the insulator-to-metal Mott transition, i.e. a transition driven by the weakening of the electronic interactions without any symmetry breaking. It has been demonstrated \cite{Brinkman_1973,guerci2019exciton} that the Hamiltonian ($\hat{H}_{loc}$) describing the EG$\rightarrow$EHL transition in photoexcited semiconductors has a one-to-one correspondence with the repulsive Hubbard model ($\hat{H}_{U}$) in a magnetic field (see Materials and Methods):
\begin{equation}
   \hat{H}_{loc}=\hat{H}_{U} - h\sum_i (n_{i\uparrow}-n_{i\downarrow})
   \label{Hubbard}
\end{equation}
where $U$ is the local electronic Coulomb repulsion, $h$ is the effective magnetic field that controls the electronic occupation $n$ at site $i$ of electrons with spin $\uparrow$ and $\downarrow$. The one-to-one correspondence between the spins in a magnetized Hubbard model and the photoexcited excitons in individual lead halide perovskite nanocubes will be discussed in detail in Sec. \ref{sec:Mott}. 

The emergence of long-range orders from incoherent fluctuations controlled by $\hat{H}_{int}$ and the Mott transition described by $\hat{H}_{loc}$ have been so far addressed separately in different materials. The lack of a single platform combining the two physics has been the main obstacle for the development of artificial simulators for real materials. Here we show that halide perovskite superlattices fill this gap by simultaneously realizing the main features that ubiquitously pervade the phase diagram of many quantum materials.

\begin{figure*}[t]
\includegraphics[width=14cm]{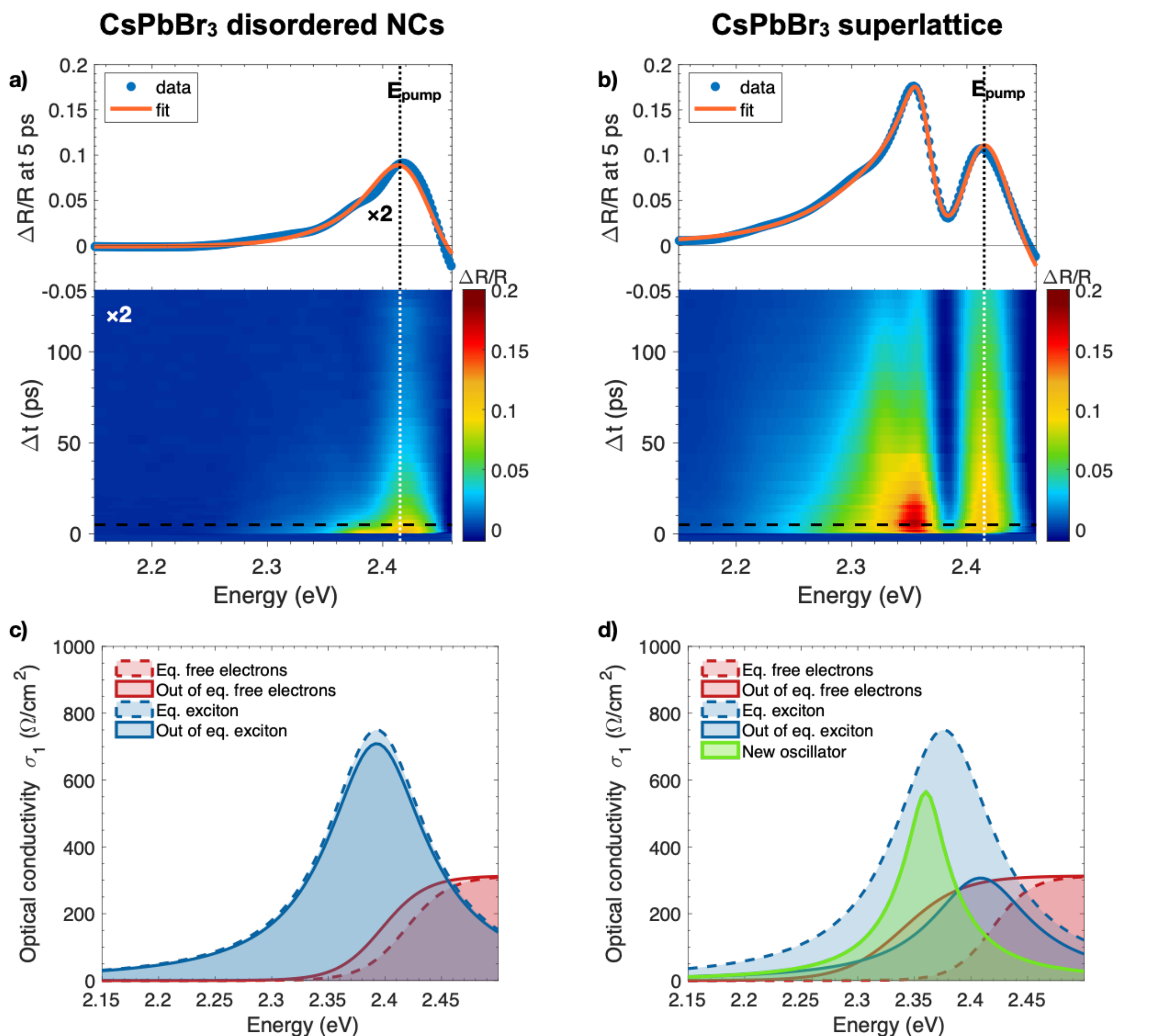}
\caption{Ultrafast transient reflectivity of \ch{CsPbBr_3} disordered NCs sample (left panels) and NC superlattice sample (right panels), measured at 17 K, 230 \textmu J/cm$^2$ excitation fluence and pump photon energy resonant with the excitonic line, i.e. $E_{pump}$ = 2.41 eV.  The bottom a) and b) panels report the two-dimensional pump-probe maps displaying the $\Delta R/R$ signal (see color scale on the right) as a function of the delay ($\Delta t$) and probe photon energy.
The top a) and b) panels report the $\Delta R/R$ signal (blue dots) as a function of the probe photon energy at fixed delay time, $\Delta t$=5 ps (horizontal dashed line in the color maps). The orange solid lines represent the differential fit to the data. c) and d): optical conductivity ($\sigma_1$) at equilibrium (dashed lines) and out of equilibrium at $\Delta t$=5 ps (solid lines) obtained from experimental absorbance and fit of $\Delta R/R$ spectra. The colors represent the different contributions to the total optical conductivity: i) main excitonic line (blue); ii) across gap optical transitions (red) and the photo-induced peak emerging at low temperature in ordered NC superlattices (green).}
\label{fig2: data}
\end{figure*}

\section{Time-resolved optical spectroscopy}
\label{TR_spectroscopy}
The physics generated by the Hamiltonian introduced in section \ref{sec: hamiltonians} is here investigated by broadband transient reflectivity measurements.
We performed  experiments on artificial lattices constituted by $L$ = 8 nm \ch{CsPbBr_3} nanocubes (Bohr exciton diameter $\sim$7 nm \cite{protesescu2015nanocrystals}) arranged in cubic superlattices of periodicity $\Lambda = L + l =$ 11 nm, where $l$ is the thickness of the ligand layer in between two neighboring NCs. The size of each superlattice is of the order of a few micrometers (1-10 \textmu m). Ultrashort light pulses are used to impulsively inject optical excitons, whose density is controlled by the light intensity. The broadband probe (2.1-2.5 eV photon energy) measures the femto/picosecond time evolution of the optical properties following the impulsive excitation. In particular, we employ a resonant pumping scheme in which the pump photon energy ($\simeq$2.41 eV) is tuned to the exciton energy, thus limiting the direct generation of free carriers in the conduction band.

 \begin{figure*}[t]
\includegraphics[width=18cm]{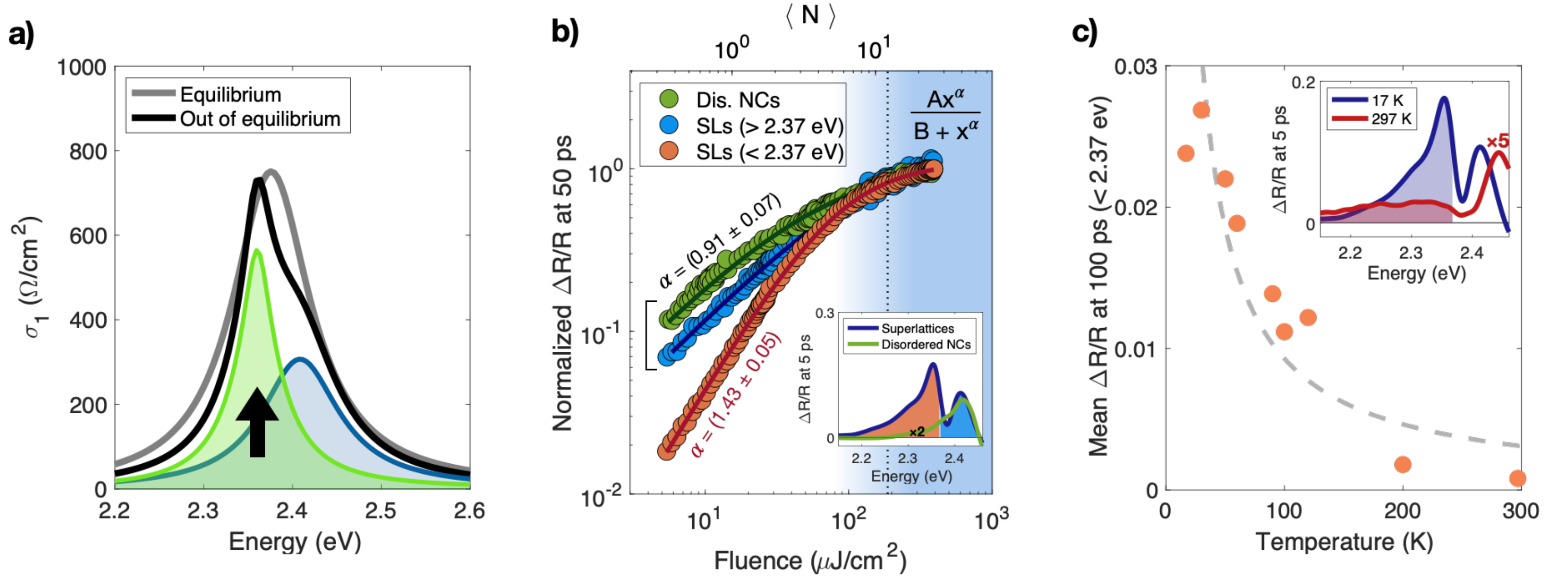}
\caption{ a) Equilibrium (gray line) and out of equilibrium (black line) excitonic resonance in \ch{CsPbBr_3} superlattices obtained from differential fit of pump-probe signal at 5 ps delay time. After pump excitation two components are distinguished: the main perovskite exciton (blue area), which undergoes a pronounced decrease of spectral weight, and the additional oscillator (green area) associated to cooperative photo-induced effects. b) Normalized amplitude of $\Delta R/R$ signal at 50 ps delay time as function of excitation intensity, represented by pump fluence on the bottom axis and  mean number of excitons per NC, $\langle N \rangle$, on the top axis. For NC superlattice samples, the data are collected by integrating the signal in the spectral region below and above 2.37 eV, as highlighted by the coloured areas in the inset (orange and blue respectively). In the top right corner we display the fitting function, where x represents both fluence and $\langle N \rangle$. The dotted vertical line represents the excitation density threshold for the Mott transition, as determined from the temporal dynamics discussed in Sec. \ref{sec:Mott}. c) Temperature dependence of $\Delta R/R$ at 100 ps delay time, integrated in the spectral region highlighted in the inset ($\hbar \omega <$ 2.37 eV). Dashed gray line: guide to the eye showing the typical $T^{-1}$ dependence of coherence phenomena.}
\label{fig3: superradiance}
\end{figure*}

Figures \ref{fig2: data}a and b display the typical data collected from pump-probe experiments (see Materials and Methods for experimental details) on \ch{CsPbBr_3} disordered NCs and NC superlattices, respectively. The samples are cooled down to 17 K and are excited by 230 \textmu J/cm$^2$ fluence pulses. The bottom panels show the reflectivity variation $\Delta R/R$ induced by the pump pulse, as a function of time delay ($\Delta t$) and probe photon energy. The top panels report a horizontal cut of the data, representing the spectrally resolved reflectivity variation of the system 5 ps after the pump excitation. For the disordered sample (see Fig. \ref{fig2: data}a), the signal is characterized by a positive reflectivity variation of the order of 5\% centered at 2.41 eV. The same experiment performed on NC superlattices (see Fig. \ref{fig2: data}b) displays a larger signal amplitude, with a similar spectral response around 2.41 eV and, additionally, a more structured spectral response extending down to $\simeq$2.20 eV probe energy.\\
The origin of these structures in the $\Delta R/R$ signal is assessed by performing a differential fit, which consists in modifying the parameters of the model describing the equilibrium optical properties that are responsible for the observed reflectivity variation. The starting point is the equilibrium optical conductivity that is obtained from a Kramers-Kr\"{o}nig constrained model matching the experimental absorbance of the samples and the temperature dependent trends reported in literature (Supplemental Material S4).
In the 2-2.5 eV energy range, the model is dominated by the conduction band edge absorption (red dashed line in Fig. \ref{fig2: data}c) and the exciton peak (blue dashed line), modelled though a Drude-Lorentz oscillator. The outcome of the differential fitting procedure is depicted in Figures \ref{fig2: data}c and d, where the solid lines represent the out of equilibrium components of the optical conductivity $\sigma_1$ necessary to fit the experimental $\Delta R/R$ signal (see orange solid lines in the top panels of Fig. \ref{fig2: data}a,b and Supplemental Material S5 for details about the robustness of the fitting procedure). In order to reproduce the measured spectral response of \ch{CsPbBr_3}, for both samples it is necessary to assume: i) a decrease of the excitonic spectral weight and a concomitant blueshift of the exciton energy, and ii) an increase of in-gap free-electron states accounted for by a red-shift of the semiconducting band gap. In addition to the i,ii) contributions observed in both samples, the feature observed in the 2.2-2.4 eV energy range solely for NC superlattices (see Fig. \ref{fig2: data}b) requires an additional narrow structure (iii) that we model through a new oscillator (green line in Fig. \ref{fig2: data}d) appearing in the out of equilibrium optical conductivity.

In the next sections we will discuss separately the different contributions to the out-of-equilibrium reflectivity. Specifically, in Sec. \ref{sec:SR} we will show that the new spectral feature iii) scales superlinearly with the excitation intensity, and disappears at $\simeq$200 K and in disordered samples, thus constituting the signature of a cooperative superradiant dynamics. In Sec. \ref{sec:Mott}) we will demonstrate that the photo-induced increase of the spectral weight of in-gap free electrons states (ii) indicates the transition from an insulating excitonic Mott phase to an electron-hole liquid phase. In Sec. \ref{sec:phasediagram}  we will combine the different time and fluence-dependent results to trace a trajectory in the phase diagram of $\hat{H}_{int}$ and $\hat{H}_{loc}$ and demonstrate that the parameters spanned by time-resolved experiments are relevant for real materials.

\section{Cooperative superradiant effects} 
\label{sec:SR}

Figure \ref{fig3: superradiance}a reports the details of the photo-induced changes in the excitonic resonance observed for NC superlattices at 5 ps delay time. The grey and black lines represent the exciton contribution to the equilibrium and out of equilibrium optical conductivity, respectively, as obtained from the fit described above. After the interaction with the excitation light pulse, we observe a $\simeq$60\% spectral weight decrease of the excitonic peak, whose out of equilibrium optical conductivity is shown by the blue filled area in Fig. \ref{fig3: superradiance}a, along with the appearance of a new resonance, represented by the green area and highlighted by the black arrow. The linewidth of the new peak is $\simeq$ 40 meV, to be compared to 100 meV width of the main excitonic line. This second component is centered at 2.36 eV, corresponding to a red-shift $\delta$ with respect to the instantaneous position of the main excitonic resonance.  By comparing results obtained on 9 different NC superlattice samples with the same nominal characteristics, we obtain a value of $\delta$ that varies between 40 and 80 meV. The estimated $\delta$ allows us to exclude multi-excitons effects, which would cause smaller (10-30 meV) red-shifts in similar systems, \cite{raino2016single,fu2017neutral,aneesh2017ultrafast,tang2022electronic} and would appear also in disordered NCs. Sample to sample variability manifests also in the fine structure of the $\Delta R/R$ signal for $\hbar \omega < $ 2.37 eV. As discussed in the Supplemental Material (Sec. S5), in some samples two distinguished structures (e.g. see data in fig. \ref{fig2: data}b at long $\Delta t$) are visible particularly for long delays. The presence of these multiple fine structures, which share the same origin and characteristics, is related to local inhomogeneity in the NCs and superlattice sizes. Samples of various aging have been measured and always show a similar response, with variability smaller than that observed from different points on the same sample.

The narrow additional peak emerging in the transient reflectivity properties is visible for both high excitation fluence (230 \textmu J/cm$^2$, Figure \ref{fig2: data}b,d) and low pump intensity (see Fig. S12 corresponding to 30 \textmu J/cm$^2$ pump fluence) and features characteristics very similar to the superfluorescence recently observed in perovskite superlattices by means of low-temperature photo-luminescence \cite{raino2018superfluorescence,findik2021high}. 
In order to unambiguously address the cooperative origin of this emergent spectral feature, we performed a detailed fluence dependence study. In Figure \ref{fig3: superradiance}b we plot the $\Delta R/R$ signal at a fixed time delay (50 ps) as a function of the excitation intensity and integrated over selected spectral regions of interest, which are highlighted by the filled areas in the inset. The top x-axis in Figure \ref{fig3: superradiance}b reports the excitation intensity expressed as the mean number of excitons in each NC, $\langle N \rangle$ (estimated as described in Section S7). The signal amplitude tends to saturate when $\langle N \rangle \gg$1, in accordance with what is reported in literature for similar systems \cite{aneesh2017ultrafast,gramlich2020thickness,klimov2000optical}. This behavior can be well described, as reported in Ref.  \citenum{klimov2000optical}, by an empirical function of the form $a\langle N \rangle^{\alpha} /(b + \langle N \rangle^{\alpha} )$, which accounts for both a low-fluence power-law increase and the high-fluence saturation. If we consider the integrated signal for $\hbar \omega>$ 2.37 eV, i.e. far from the photoinduced additional peak, we obtain ${\alpha} = (0.91 \pm 0.07)$, which corresponds to a linear behaviour at low fluence. In contrast, the spectral region corresponding to the additional photo-induced peak ($\hbar \omega<$ 2.37 eV) features a clearly superlinear fluence dependence, corresponding to ${\alpha} = (1.43 \pm 0.05)$, in agreement with what expected for superradiant phenomena \cite{raino2018superfluorescence,biliroglu2022room,cherniukh2021perovskite,krieg2020monodisperse}. It is useful to compare the present results with those obtained on disordered NCs, in which collective superradiant phenomena should be quenched \cite{Masson2020,Masson2022,Sierra2022} due to disorder-driven dephasing. As reported in Fig. \ref{fig3: superradiance}b (green dots and lines), the fluence-dependence of the signal is always linear, independent of the energy region considered.
We note that the fluence dependent signals shown in Fig. \ref{fig3: superradiance}b collapse to the same curve above a fluence of approximately  150-200 \textmu J/cm$^2$. This threshold value, which we will indicate as $F^t$ in the following, indicates that the observed cooperative effect is sustained only for moderate densities of excitons, as it will be discussed in more detail in the next sections. 

The cooperative origin of the photo-induced structure at 2.36 eV observed on the ordered NC superlattices is further corroborated by the temperature dependence data reported in Figure \ref{fig3: superradiance}c. The top-right panel shows the transient reflectivity spectra 5 ps after pump excitation at 17 K (blue) and 300 K (red), the latter being multiplied by a normalization factor for comparison purposes. At room temperature, the reflectivity variation corresponding to the spectral feature at 2.36 eV is considerably reduced. In the main graph we report the $\Delta R/R$ signal at 100 ps, integrated in the 2.15-2.37 eV energy region as a function of temperature $T$. The suppression of the photo-induced peak at temperatures as high as 200 K is compatible with the thermally driven loss of coherence of superradiant emitters  \cite{mattiotti2020thermal}. The comparison to a $T^{-1}$ guide to the eye (gray dashed line in Fig. \ref{fig3: superradiance}c) indicates a fair agreement with the scaling behavior of the superradiance enhancement factor, that reduces to 1 in the high temperature limit according to a $T^{-1}$ power law \cite{mattiotti2020thermal}.

Lastly, we note that, in the same spectral range (2.30-2.37 eV), intense fluorescence induced by the pump beam is observed under the form of stray light emitted by NC superlattices and arriving at the detector (see Figure S17). This fluorescence is suppressed by $\sim$70\% in the disordered NCs, thus further demonstrating that the superlattice is the key element for achieving the collective superradiant regime described by the inter-unit cell hamiltonian $\hat{H}_{int}$ (see Materials and Methods).

\section{the Mott transition}
\label{sec:Mott}
\begin{figure}
\includegraphics[width=8.5 cm]{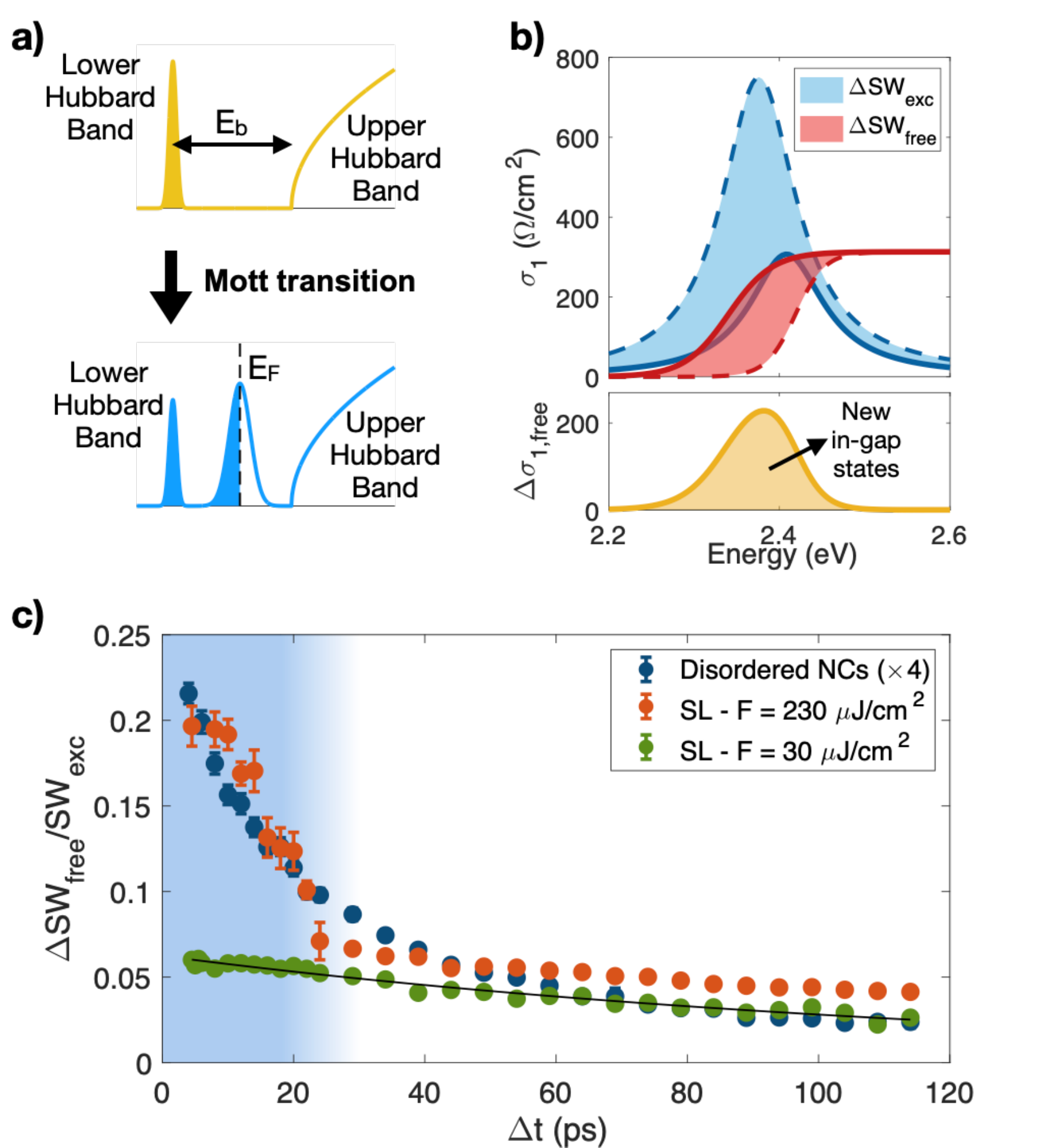}
\caption{a) Sketch of the Mott transition from an insulator to a metal where new metallic states appear at the Fermi level. b) Top panel: plot of the equilibrium (dashed lines) and out of equilibrium (solid lines) optical conductivity obtained from fitting the experimental data at $\Delta t$ = 5 ps. The blue filled area represents the excitonic spectral weight decrease, which is counterbalanced by the appearance of new states below the conduction band minimum, described by a red-shift of the band edge (red filled area). Bottom panel: difference between out of equilibrium and equilibrium $\sigma_{1,free}$, which  represents the contribution from the conduction band states to the optical conductivity. c) Time evolution of the spectral weight transfer from the exciton state to free carrier states, estimated as $\Delta SW_{free}/SW_{exc}$. The black line denotes an exponentially decaying function fitted to the data. The blue area highlights the region where an excess of $\Delta SW_{free}$ is observed for large excitation fluence.}
\label{fig4: Mott}
\end{figure}

The transition from EG to EHL that takes place in photoexcited semiconductors has been thoroughly discussed in Ref. \citenum{guerci2019exciton}. This problem has a one-to-one correspondence with the Mott transition in a magnetized Hubbard model, described by $\hat{H}_{loc}$, in which the electron spins play the role of the electron-hole excitations. In this effective description, the density of photoinduced excitons, $n_{\mathrm{eh}}$, is mapped into the number of flipped spins with respect to the ferromagnetic background, i.e. $n_{\mathrm{eh}}$=$\sum_i \langle n_{i\downarrow} \rangle/N$ ($N$ being the total number of particles). The exciton density is therefore controlled via the auxiliary magnetic field $h$ that induces an effective magnetization $m$=1-2$n_{\mathrm{eh}}$. For large $h$, the system is fully polarized ($m$=1) and no excitons are present. The sudden reduction of $h$, which injects a finite numbers of spin excitations ($m<$1), mimics the sudden photoinjection of excitons ($n_{\mathrm{eh}}>$0). The other parameter controlling the Mott transition is the repulsive Coulomb interaction $U$, which determines the excitonic binding energy $E_{b}$. At large $U$ and $m$, the magnetized Hubbard model is characterized by insulating solutions constituted by localized spins that are frozen by the electronic repulsion. When either $m$ or $U$ decrease, the system undergoes a transition towards a metallic state in which the spins are delocalized and can hop throughout the lattice. Interestingly, in a large parameter phase space the transition is of first order nature and it is therefore characterized by real space coexistence of insulating and metallic solutions with different spin densities \cite{guerci2019exciton}. In terms of the excitonic problem, the insulating phase of the magnetized Hubbard models maps the existence of well defined localized excitonic states (lower Hubbard band), separated from the upper Hubband band by an energy amount $E_{b}$ (see Fig. \ref{fig4: Mott}a). When the excitonic density $n_{\mathrm{eh}}$ is increased to values of the order of 1-10\%, the emergence of metallic states, corresponding to a liquid of delocalized electrons and holes, manifests itself in the appearance of new in-gap states at the expenses of the excitonic states (see Fig. \ref{fig4: Mott}a). 

Here, this physics is accessed in the saturation region (see Fig. \ref{fig3: superradiance}b) at excitation intensities of the order of $\sim$200 \textmu J/cm$^2$. As shown in Sec. \ref{TR_spectroscopy}, in this regime the transient optical response of NC superlattices is dominated by a decrease of the excitonic spectral weight, and the corresponding increase of in-gap free electrons states, as indicated by the effective red-shift of the semiconducting gap (see Figs. \ref{fig2: data}d and \ref{fig4: Mott}b). Interestingly, this kind of response, surviving on a $\sim$ 100 ps timescale, is remarkably different from what usually reported for above-resonance excitation experiments in similar materials, where the large number of free carriers injected in the conduction band leads to a very fast band-gap renormalization followed by a band filling effect after $\sim$ 1 ps \cite{aneesh2017ultrafast,franceschini2020tuning,price2015hot}. The transient increase of in-gap states at the expenses of the intensity of the exciton peak suggests that, at high fluence, the NC superlattices no longer support well defined excitons but rather delocalized electron-hole excitations.

To assess the nature of this high-excitation regime, we calculated from the differential model (see Supplemental Material S5) the total spectral weight:
\begin{equation}
    SW_{tot} = \int_0^{+\infty} \sigma_{1,tot}(\omega) d\omega 
\end{equation}
which is, by definition, a conserved quantity depending on the total number of electrons in the system \cite{wooten1972}. In particular, we obtained from the data the different contributions:
\begin{equation}
    SW_{tot} = SW_{2.36 eV}+SW_{exc}+SW_{free}
\end{equation}
where $SW_{2.36 eV}$ is the spectral weight of the photo-induced cooperative peak at 2.36 eV, $SW_{exc}$ is the spectral weight of the excitonic peak already present in the equilibrium optical conductivity, and $SW_{free}$ is the spectral weight associated to the direct across-gap transitions. The analysis indicates that the photo-induced decrease of the excitonic spectral weight ($\Delta SW_{exc}\simeq$ -74 $\Omega$eV/cm$^2$ at $\Delta t$ = 5 ps) is perfectly compensated by both the new peak at 2.36 eV ($\Delta SW_{2.36 eV}\simeq$ 49 $\Omega$eV/cm$^2$) and by an increase of in-gap free electron states ($\Delta SW_{free}\simeq$ 25 $\Omega$eV/cm$^2$). Normalization to the total spectral weight of the excitonic peak returns $\Delta SW_{free}/SW_{exc}\simeq$ 0.2, indicating that approximately 20\% of the initial exciton SW is transferred into in-gap free electrons states. We note that, in the saturation regime, the number of excitons for each single nanocube is of the order of $\langle N \rangle$=20 (see Fig. \ref{fig3: superradiance}b), which corresponds to a photodoping $n_{\mathrm{eh}}\sim$1\% (see Sec. S7). This value cannot account for the observed spectral weight change, unless we assume a transformation of the electronic band structure of the system. The spectral weight transfer from excitonic states to free-electron ones can be therefore considered as the direct manifestation of the excitonic Mott transition \cite{guerci2019exciton}, as captured by $\hat{H}_{loc}$ (see Eq. \ref{Hubbard} and Materials and Methods Eq. \ref{eq: Hubbard}). 

The time-resolved dynamics contains important information about the temporal evolution of the newly created metallic states and the recovery of the initial excitonic gas. Since during the relaxation $n_{\mathrm{eh}}$ decreases due to the slow recombination across the semiconducting gap, at some time the system will undergo  the transition from the photo-induced EHL state back to the EG insulating phase.
In Fig. \ref{fig4: Mott}c we plot the fraction of the excitonic spectral weight that is transferred to free carrier states as a function of pump-probe time delay $\Delta t$. 
At low fluence ($\sim$30 \textmu J/cm$^2$, green markers in Fig. \ref{fig4: Mott}c), the spectral weight of the photoinduced metallic states is very limited and exponentially decays with a timescale of 130 ps (black solid line in Fig. \ref{fig4: Mott}c). This slow relaxation is in agreement with what expected for the recombination of electron and holes across the gap and with the fluorescence timescale \cite{raino2018superfluorescence,diroll2018low}. At large fluences ($\sim$230 \textmu J/cm$^2$, red markers in Fig. \ref{fig4: Mott}c), we observe additional spectral weight variation which exceeds that present in the low-fluence data. This additional $\Delta SW_{free}$ component rapidly relaxes with a timescale of $\sim$20 ps, thus allowing us to estimate the critical number of excitations necessary for re-establishing the insulating EG phase. Assuming that $n_{\mathrm{eh}}$ spontaneously decays with the timescale of 130 ps, the change in slope of the $\Delta SW_{free}/SW_{exc}$ dynamics at $\sim$20 ps corresponds to a threshold value ${n}^t_{\mathrm{eh}}$=0.5\%. If we calculate the pump fluence necessary to inject this density of photoexcitations, we obtain a threshold fluence $F^t\simeq$190 \textmu J/cm$^2$, which is compatible with the fluence at which cooperative effects saturate (see Sec. \ref{sec:SR}). A similar conclusion is obtained by analyzing the data on disordered NCs, which display a similar change in slope of the $\Delta SW_{free}/SW_{exc}$ dynamics (see Fig. \ref{fig4: Mott}c). We note that, although the dynamics in disordered NC is qualitatively very similar to what observed in ordered samples at high fluence, the measured signal is significantly smaller. This observation suggests that, although the Mott transition takes place in both ordered and disordered NCs, finite size effects \cite{Wang2002}, which go beyond the scope of this work, emerge when NC are arranged in a disconnected and random network (disordered NCs) in which coherent hopping and inter-cube delocalization is suppressed \cite{Blach2022}.

\section{the phase diagram}
\label{sec:phasediagram}
\begin{figure}
\includegraphics[width=8.5cm]{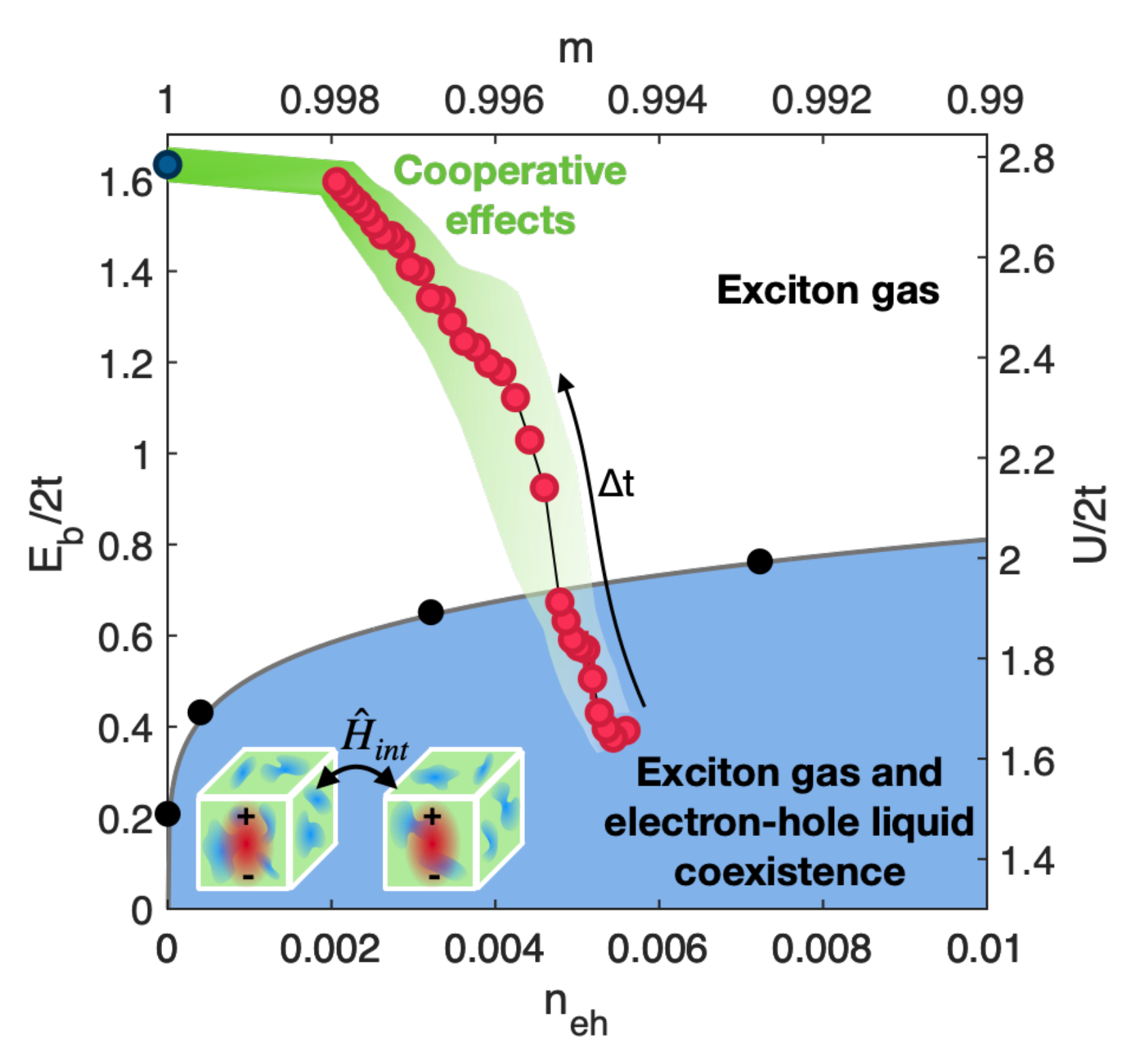}
\caption{Phase diagram showing the electron gas (EG) and electron-hole liquid (EHL) regions. The electron-hole density $n_{\mathrm{eh}}$ and the exciton binding energy $E_{b}$ correspond to the magnetization $m$ and the on-site interaction $U$ in the equivalent Hubbard model. The black points are taken from Ref. \citenum{guerci2019exciton}. The red dots are the experimental data points representing the trajectory the system follows while relaxing to the equilibrium EG phase from the photo-induced EHL phase. The plotted values are the binding energy values extracted from the time-resolved fit of the pump-probe data on NC superlattices at 230 \textmu J/cm$^2$ fluence, as a function of the estimated electron-hole density at the corresponding $\Delta t$. The green shaded area represents the phase-space region compatible with the outputs of the time-resolved experiment when the error bars associated to $n_{\mathrm{eh}}$ and $E_{b}$ are considered.}
\label{fig5: Phase diag}
\end{figure}

Establishing halide perovskites artificial solids as quantum simulators is crucially based on the possibility of controlling the Hamiltonians $\hat{H}_{loc}$ and $\hat{H}_{int}$ and spanning a phase-space region that is relevant for describing real correlated materials \cite{Georges1996,Kotliar2004}. Considering relevant examples, such as vanadium \cite{Imada1998}, manganese \cite{Salamon2001}, and copper oxides \cite{Keimer_2015,Comanac2008} or fullerides \cite{Capone2002,Capone2009}, the fundamental ground state properties can be captured by the Hubbard model $\hat{H}_{U}$, formally identical to that introduced in Eqs. \ref{Hubbard} and \ref{eq: Hubbard} (Materials and Methods). In these materials, the effective Coulomb repulsion $U$ is of the order of the bandwidth 4$t$ (with $t$ being proportional to the hopping parameter, following the notation of Ref. \citenum{guerci2019exciton}), thus leading to a Mott insulating ground state \cite{Imada1998}. When additional free carriers are introduced by chemical doping \cite{Lee2006}, these doped Mott insulators progressively develop spatially inhomogeneous correlated metallic states or undergo insulator-to-metal phase transitions \cite{Imada1998}. In many interesting cases, the doped Mott insulating phase witnesses the emergence of low-temperature long-range collective states \cite{Lee2006} such as unconventional superconductivity and spin- or charge-density waves. Independently of the specific microscopic mechanism (e.g. electron-phonon or electron-spin interactions) at play in different materials, the formation of collective macroscopic states is mediated by long-range interactions, similar to those contained in $\hat{H}_{int}$, that favour the phase-locking of fundamental incoherent local fluctuations.

In this section we discuss how the fluence and time-dependent experiments presented in the previous sections allow to access an important region of the zero-temperature phase diagram of $\hat{H}_{loc}$+$\hat{H}_{int}$ \cite{guerci2019exciton}. More specifically, the parameters controlling the electronic phases of photoexcited halide perovskites are the excitation density $n_{\mathrm{eh}}$, which is mapped into the magnetization $m$ through $\hat{H}_{loc}$, and the binding energy $E_b$, which is mapped into the Coulomb repulsion $U$ (see Sec. \ref{sec:Mott}). For small excitation densities ($n_{\mathrm{eh}}<$1\%) and moderately large Coulomb repulsion ($U>3.8t$), the phase diagram is characterized by the boundary between an excitonic insulating gas and a mixed state with phase separation between EG and EHL, as shown in Fig. \ref{fig5: Phase diag} \cite{guerci2019exciton}. The resonant excitation of the excitonic line directly modifies $n_{\mathrm{eh}}$ without creating an additional effective electron-hole population which would require a finite temperature description. At the same time, as anticipated in Sec. \ref{TR_spectroscopy}, the excitonic energy undergoes a transient blueshift, $\delta E_b$ (see Fig. \ref{fig2: data} and related discussion), that is maximum at short delays ($\sim$ 5 ps) and progressively decreases as the system relaxes and returns to the initial state. The observed blue shift is likely related to a dynamical weakening of the exciton binding energy as a consequence of the increased screening after the light excitation. The determined time-dependent values of $n_{\mathrm{eh}}$ and $\delta E_b$ define a trajectory in the phase diagram, in which the energy scales are expressed as a function of the unknown effective hopping $t$. The determination of ${n}^t_{\mathrm{eh}}$=0.5\% as the density threshold for the instability of the EG (see Sec. \ref{sec:Mott}), allows us to anchor the time-dependent trajectory and fix the range of the $U/2t$ values spanned by the time-resolved experiment. Fig. \ref{fig5: Phase diag} shows the trajectory in the phase diagram for the high-fluence experiment. At very short delays, the NC superlattices are driven into a non-equilibrium state corresponding to $U/2t\approx$1.6 and $n_{\mathrm{eh}}$=0.6\%, which is characterized by phase separation between insulating EG and metallic EHL regions. In this regime, cooperative phenomena start to be quenched (see Fig. \ref{fig3: superradiance}b) due to the progressive growth of metallic domains. During the relaxation dynamics, the system undergoes a dynamical transition back to the EG insulating phase before the initial parameters ($U/2t\approx$2.8 and $n_{\mathrm{eh}}$=0) are recovered on longer timescales.

When combined together, our results demonstrate the possibility to explore the region of the magnetized Hubbard model $\hat{H}_{loc}$ that is the most relevant to describe many-body effects in correlated materials ($U$=3.2-5.6$t$).
Importantly, perovskite NC superlattices also host long-range collective states (superradiance) driven by $\hat{H}_{int}$ (see Sec. \ref{sec:SR}) in the proximity of the insulator-to-metal transition controlled by $\hat{H}_{loc}$, i.e. for values of $U$ ranging from $\approx$4$t$ to $\approx$5.6$t$ (see Fig. \ref{fig5: Phase diag}). Although in the present case we access the insulator-to-metal transition in $\hat{H}_{loc}$ only in the presence of a magnetic field $h$,  the observed phenomenology is a very general property of correlated materials, such as iron-based superconductors, superconducting copper oxides and fullerides, which develop superconductivity and other long-range collective phases, e.g. charge density waves, nematicity, antiferromagnetism, spin density waves, when the Mott insulating state is lightly doped \cite{Keimer_2015,Keimer_2017}.

\section{conclusions and outlook}

In conclusion, we have demonstrated that perovskite NC artificial solids represent a novel platform to investigate the emergence of long-range cooperative phases in systems displaying a Mott insulator-to-metal transition. The possibility to simulate the physics of interacting systems in which local correlations and long range orders interact on similar time (5-100 ps) and spatial ($\Lambda \simeq$10$a$) scales has direct impact on a more general class of solid-state problems. The most relevant case is the physics of copper oxides that host unconventional superconductivity and other exotic orders when the insulating Mott state is properly doped \cite{Keimer_2015,Keimer_2017}. Controlling the interplay between short-range Coulomb interactions and macroscopic collective states \cite{Peli2017} is key to understand how phase-coherence on length scales larger than the lattice spacing may emerge out of incoherent fluctuations of pre-formed Cooper pairs or of the charge distribution and give rise to the superconducting condensate and charge density waves.
The full tunability of $\Lambda$, of the exciton energy and density, allows to artificially reproduce the phase diagram of cuprates or other quantum materials and implement models that describe the formation of long-range collective phase through the interaction term $\hat{H}_{int}$. 

We also foresee that time-resolved experiments will provide tools to unlock the gate to wider region of the phase diagram and a richer physics.
Non-resonant experiments will allow to directly create a non-thermal electron-hole population that constitutes a thermal reservoir at a very high effective temperature, thus providing a platform with controllable disorder and lattice size to simulate the magnetized Hubbard model at finite temperatures.

\subsection*{Acknowledgments}
C.G., M.C., P.F., A.M., S.M. acknowledge financial support from MIUR through the PRIN 2015 (Prot. 2015C5SEJJ001) and PRIN 2017 (Prot. 20172H2SC4\_005) programs. C.G., S.P. and G.F. acknowledge support from Universit\`a Cattolica del Sacro Cuore through D.1, D.2.2 and D.3.1 grants. S.M. acknowledges partial financial support through the grant "Finanziamenti ponte per bandi esterni" from Universit\`a Cattolica del Sacro Cuore. 

\section*{Materials and methods}
\textbf{Model Hamiltonians} The long-range interaction among excitons in different quantum dots is driven by the transverse electromagnetic field and can be expressed as
\begin{equation}
\hat{H}^{(\mathbf{K})}_{int}\propto(a+a^\dagger)\sum_{\bm{K}} g_{\bm{K}} [d^{\dagger}_{\bm{K}c} d_{\bm{K}v}+d^{\dagger}_{\bm{K}v} d_{\bm{K}c}]
    \label{eq: dipole}
\end{equation}
where $g_{\bm{K}}$ is the dipole transition element and $\bm{K}$ is a wavevector in the first Brillouin zone of the superlattice. $d^{\dagger}_{\bm{K}v}$ and $d^{\dagger}_{\bm{K}c}$ ($d_{\bm{K}v}$ and $d_{\bm{K}c}$) are the creation (annihilation) operators for an electron with momentum $\bm{K}$ in the valence and conduction band, respectively.  The terms $a$ and $a^\dagger$ denote the bosonic creation and annihilation operators for the photon field. 

The local Hamiltonian $\hat{H}_{loc}$ can be mapped onto a spin-polarized repulsive Hubbard model \cite{guerci2019exciton}:
\begin{widetext}
\begin{equation}
\label{eq: Hubbard}
   \hat{H}_{loc} = -t \sum_{\langle ij\rangle,\sigma} (c_{i,\sigma}^\dagger c_{j,\sigma}+c_{j,\sigma}^\dagger c_{i,\sigma})
    + U \sum_{i} n_{i\uparrow}n_{i\downarrow} - h\sum_i (n_{i\uparrow}-n_{i\downarrow})
\end{equation}
\end{widetext}
where $t$ is the hopping parameter, $c_{i,\sigma}^\dagger$ and $c_{i,\sigma}$ are creation and annihilation operators for an electron of spin $\sigma$ at site $i$, $n_{i\sigma}$ is the fermionic occupation number, $U$ is the on-site repulsion, and $h$ is the magnetic field.\\

\textbf{Samples.} \ch{CsPbBr_3} superlattices are cubic arrays of \ch{CsPbBr_3} perovskite nanocrystals. The fabrication of superlattices is accomplished by drop-casting a toluene dispersion of nanocubes of mean lateral size $\sim$ 8 nm onto a silicon substrate and letting the solvent  evaporate overnight. During this process, the nanocubes spontaneously assemble into superlattices with sizes of 1-10 \textmu m. Nanocrystal synthesis and superlattice preparation are further detailed in the Supplemental Material. Figure S5 displays optical microscope and SEM images of \ch{CsPbBr_3} nanocube superlattices on glass showing their morphology. XRD data characterising the packing of nanocubes is reported in Fig. S3b. \\ 
Disordered nanocubes films are fabricated starting from the same synthetic batches of superlattice films, and using fast solvent evaporation and mechanical scrambling (e.g. spreading or crushing). The disordered nature of nanocubes films prepared in such a way was characterized by X-ray diffraction (see figure S3a of the Supplemental Material). The lack of preferred orientation and superlattice diffraction signatures in diffraction patterns \cite{toso2019wide,toso2021multilayer} was taken as an evidence of random orientations of nanocubes in the sample (see Supplemental Material). \\
The optical properties of the samples are characterized by room temperature absorbance and photoluminescence, reported in figures S2a and S2b for, respectively, disordered NCs and \ch{CsPbBr_3} superlattices  deposited on a transparent substrate.\\
The time-resolved spectroscopic measurements were performed on superlattice and disordered film samples of various age. The morphological and structural characterization of the aged samples as compared to freshly-prepared ones are documented in the Supplemental Material.\\ 

\textbf{Pump-probe spectroscopy.} The out of equilibrium properties of \ch{CsPbBr_3} superlattices and disordered NCs samples are investigated by means of broadband transient reflectivity experiments. The excitation is a 250 fs laser pulse at 2.41 eV photon energy (resonant with the excitonic level of the perovskite compound) and it is obtained by frequency doubling the emission of the laser system (Pharos by Light Conversion) in a 1-mm-thick BBO crystal. The induced variations in reflectivity are probed with a supercontinuum pulse, produced by means of White Light Generation process in a 4-mm-thick sapphire crystal pumped by a 1.6 eV photon energy pulse. The time delay between the arrival times of pump and probe pulses is controlled through a linearly motorized stage which delays the pump pulse in a time window covering $\sim$ 300 ps. The pump beam is focused to a 200 \textmu m $\times$ 300 \textmu m spot size, being $\approx$ 10 times larger than the probe spot size at the sample position (23 \textmu m $\times$ 23 \textmu m). The excitation intensity can be continuously varied between 0 and 400 \textmu J/cm$^2$ by rotating a half-waveplate positioned on the pump beam path and followed by a polarizer that transmits the horizontally polarized component of the light. The probe beam is vertically polarized in order to allow filtering of the signal background, which is mainly given by sample scattering of the pump beam. The laser repetition rate employed in the measurements presented here is 400 kHz; no change in the sample response is observed upon decreasing the repetition rate while keeping fixed the energy per pulse of pump and probe beams. The signal detection is performed by lock-in acquisition of the reflected probe interferogram (generated by GEMINI interferometer, NIREOS) and computation of its Fourier transform at each fixed pump-probe time domain \cite{preda2016broadband}. For pump-probe measurements, the samples are mounted inside a closed-cycle helium cryostat that allows to perform the ultrafast optical spectroscopy experiments at temperatures between 17 K and 300 K. \\

\newpage

\bibliography{Refs}

\end{document}


\title{Supplementary information for\\ Halide perovskite artificial solids as a new platform to simulate collective phenomena in doped Mott insulators}

\author{Alessandra Milloch}
\email[]{alessandra.milloch@unicatt.it}
\affiliation{Department of Mathematics and Physics, Università Cattolica del Sacro Cuore, Brescia I-25133, Italy}
\affiliation{ILAMP (Interdisciplinary Laboratories for Advanced
Materials Physics), Università Cattolica del Sacro Cuore, Brescia I-25133, Italy}
\affiliation{Department of Physics and Astronomy, KU Leuven, B-3001 Leuven, Belgium}

\author{Umberto Filippi}
\affiliation{Italian Institute of Technology (IIT), Genova 16163, Italy}

\author{Paolo Franceschini}
\affiliation{CNR-INO (National Institute of Optics), via Branze 45, 25123 Brescia, Italy}

\author{Michele Galvani}
\affiliation{Department of Mathematics and Physics, Università Cattolica del Sacro Cuore, Brescia I-25133, Italy}

\author{Selene Mor}
\affiliation{Department of Mathematics and Physics, Università Cattolica del Sacro Cuore, Brescia I-25133, Italy}
\affiliation{ILAMP (Interdisciplinary Laboratories for Advanced
Materials Physics), Università Cattolica del Sacro Cuore, Brescia I-25133, Italy}

\author{Stefania Pagliara}
\affiliation{Department of Mathematics and Physics, Università Cattolica del Sacro Cuore, Brescia I-25133, Italy}
\affiliation{ILAMP (Interdisciplinary Laboratories for Advanced
Materials Physics), Università Cattolica del Sacro Cuore, Brescia I-25133, Italy}

\author{Gabriele Ferrini}
\affiliation{Department of Mathematics and Physics, Università Cattolica del Sacro Cuore, Brescia I-25133, Italy}
\affiliation{ILAMP (Interdisciplinary Laboratories for Advanced
Materials Physics), Università Cattolica del Sacro Cuore, Brescia I-25133, Italy}

\author{Francesco Banfi}
\affiliation{FemtoNanoOptics group, Université de Lyon, CNRS, Université Claude Bernard Lyon 1, Institut Lumière Matière, F-69622 Villeurbanne, France}

\author{Massimo Capone}
\affiliation{International School for Advanced Studies (SISSA), via Bonomea 265, Trieste}

\author{Dmitry Baranov}
\affiliation{Italian Institute of Technology (IIT), Genova 16163, Italy}
\affiliation{Division of Chemical Physics, Department of Chemistry, Lund University, P.O. Box 124, SE-221 00 Lund, Sweden}

\author{Liberato Manna}
\affiliation{Italian Institute of Technology (IIT), Genova 16163, Italy}

\author{Claudio Giannetti}
\email[]{claudio.giannetti@unicatt.it}
\affiliation{Department of Mathematics and Physics, Università Cattolica del Sacro Cuore, Brescia I-25133, Italy}
\affiliation{ILAMP (Interdisciplinary Laboratories for Advanced
Materials Physics), Università Cattolica del Sacro Cuore, Brescia I-25133, Italy}
\affiliation{CNR-INO (National Institute of Optics), via Branze 45, 25123 Brescia, Italy}

\maketitle

\section{Sample preparation}
\subsection{Synthesis of cesium oleate stock solution}
The cesium oleate stock solution was prepared by loading 0.4 g of cesium carbonate (\ch{Cs_2CO_3} , 99$\%$), 1.75 ml of oleic acid (OA, 90$\%$) and 15 ml octadecene-1 (ODE, 90$\%$)  into a 40 ml vial and placing it into an aluminum block preheated at 120 °C on top of a hotplate. The mixture was stirred under N$_2$ flow and its temperature was kept between $\sim$110-115 °C until all visible solid disappeared ($\sim$60 minutes). The mixture was eventually cooled down on a room temperature hotplate under stirring.
\subsection{Synthesis of \ch{CsPbBr_3} nanocubes}
\ch{CsPbBr_3} nanocubes (NCs) were synthesized following the method reported in Ref. \cite{baranov2019investigation} and Ref. \cite{toso2019wide} with minor variations.
In a 20 ml vial, 72 mg of lead (II) bromide (\ch{PbBr_2},
$\geqslant$ 98$\%$) were combined with 5 ml of ODE, 0.5 ml of oleylamine (OLAm, 70$\%$) and 0.05 ml of OA. The vial was equipped with a magnetic stirrer and placed into an aluminum block preheated at 185 °C on top of a hotplate. As the temperature of the mixture reached 170 °C, the vial was lifted from the block and fixed with a clamp above the hotplate and as soon as it cooled down to 160 °C, 0.5 ml of the cesium oleate stock solution were swiftly injected. The reaction was quenched after $\sim$ 7 seconds with a cold water bath ($\sim$ 20 °C) under stirring. The crude solution was centrifuged for 3 minutes at 4000 rpm and the supernatant was discarded. Then the solution was centrifuged again at 4000 rpm for 3 minutes to discard any remaining liquid, with the vials in the centrifuge oriented such that the precipitate was pointing outwards. The remaining liquid was removed with a 100 $\micro$l mechanical micropipette, and the step was repeated again, this time using a cotton swab to collect the liquid. The remaining solid was dissolved in 1 ml of toluene (99.7$\%$) and filtered through 0.45 $\micro$m hydrophobic PTFE syringe filter to eliminate any residual aggregates.\\
The solution for photoluminescence (PL) and absorbance measurements was prepared by diluting 6 $\micro$l of the NC solution in a 2994 $\micro$l of toluene in a quartz cuvette.
NC batches were considered suitable for self-assembly experiments when the full width at half maximum (FWHM) of the PL spectrum was < 90 meV. Absorbance was measured through 10 mm path length, and the NC concentration was calculated from Beer's law by using absorbance at 335 nm and a published size-dependent extinction coefficient \cite{de2016highly}. Typical concentration varied in the range of $\sim$ 4-4.5 $\micro$M.
\subsection{Preparation of \ch{CsPbBr_3} nanocubes superlattice and disordered film}
\begin{figure}[h!]
     \centering
    \begin{subfigure}[t]{0.45\textwidth}
    \includegraphics[width=1\textwidth]{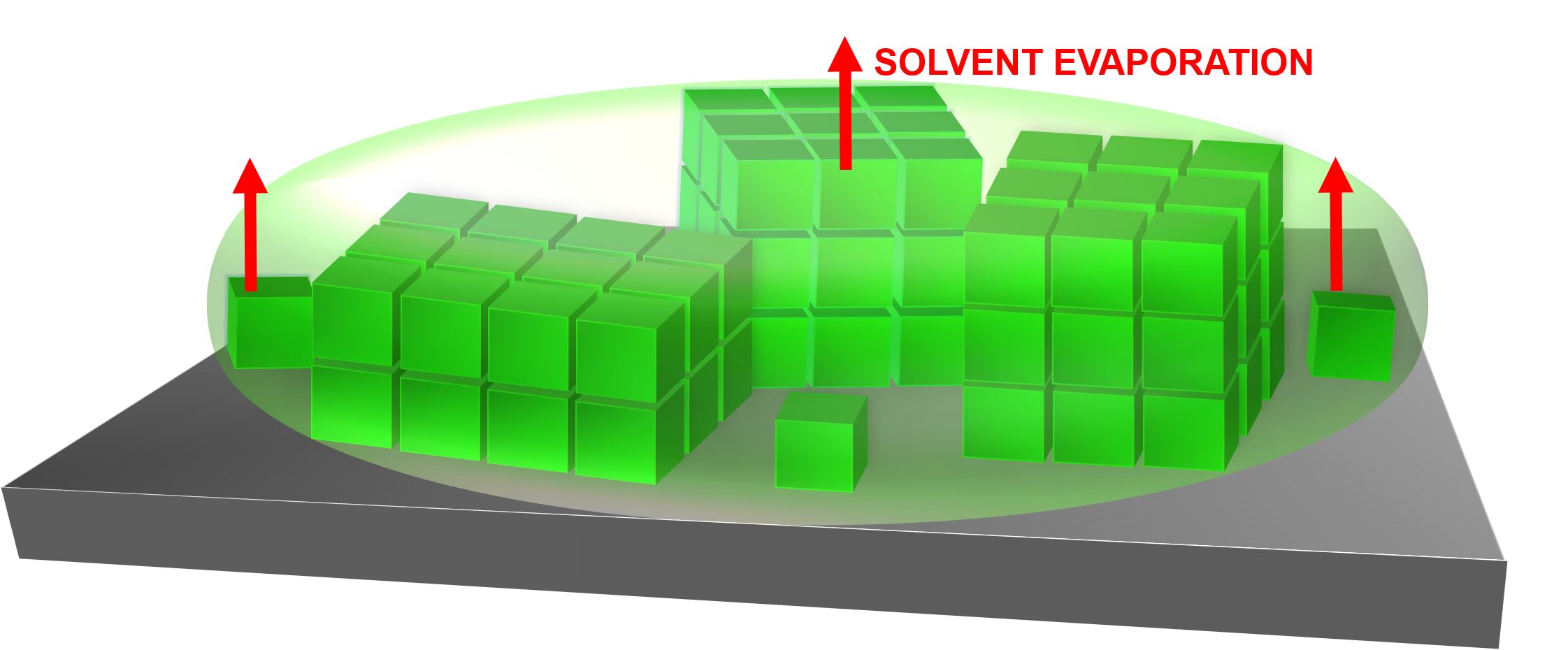}
    \caption[Specimen fixed on the paper graph mask.]{}
\label{SI - fig: sl_film}
    \end{subfigure}$\,\,\,\,\,\,\,\,\,\,\,\,\,\,\,\,\,\,\,\,\,\,\,\,$
    \begin{subfigure}[t]{0.45\textwidth}
    \includegraphics[width=1\textwidth]{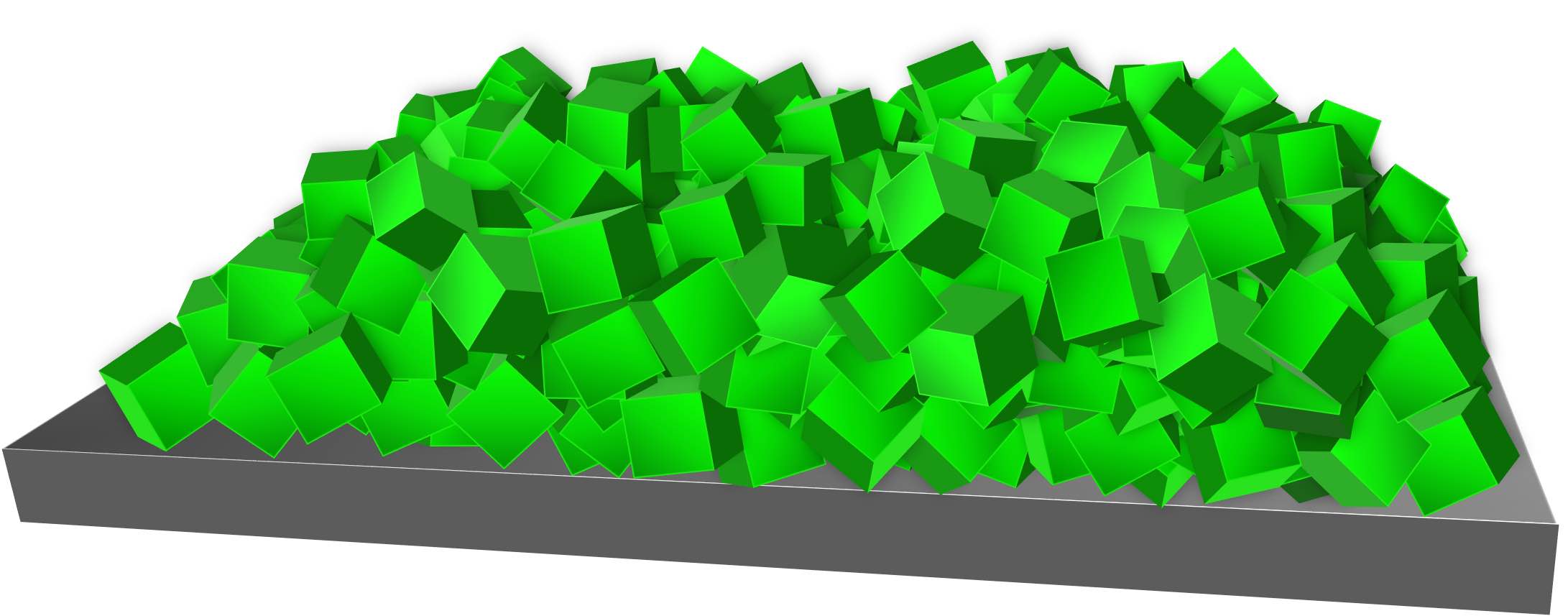}
    \caption[Specimen fixed on the paper graph mask.]{}
    \label{SI - fig: dis_film}
    \end{subfigure}
    \caption{(a) Schematic representation of SLs formation during slow solvent evaporation and of (b) randomly oriented NCs on a silicon substrate.}
    \label{SI - fig: film}
\end{figure}

Self-assembly of \ch{CsPbBr_3} NCs was carried out using a slow solvent evaporation method on 1 cm $\times$ 1 cm monocrystalline silicon (Si) and glass substrates (figure \ref{SI - fig: sl_film}). The substrates were cleaned by rinsing with methanol and 2-propanol and dried by gentle tapping with an absorbing paper tissue and by blowing compressed air. \\
Three substrates were placed inside a Petri dish and 30 $\micro$l of the stock solution of \ch{CsPbBr_3} NCs in toluene were drop-casted on each of them. The solvent was allowed to slowly evaporate ($\sim$12 hours), after which films were considered ready to be used for experiments.\\
The disordered films of \ch{CsPbBr_3} NCs were instead prepared by collecting the precipitate of the crude solution after the three centrifugation steps and by mechanically scrambling and pressing it with a plastic scoop and eventually spreading it on 0.5 cm $\times$ 1 cm Si or glass substrates (figure \ref{SI - fig: dis_film}).\\
Samples thickness was measured from SEM images of middle film profiles obtained by breaking in half replicas of samples used for experiments. Estimated sample thickness for the SL films was 7 $\micro$m $\leqslant thickness_{SL} \leqslant$ 15 $\micro$m, and for the disordered films was 10 $\micro$m $\leqslant thickness_{dis} \leqslant$ 20 $\micro$m.\\

\section{Characterization of SL and disordered films}
\subsection{Photoluminescence and absorbance}
\begin{figure}[h!]
     \centering
    \begin{subfigure}[t]{0.45\textwidth}
    \includegraphics[width=1\textwidth]{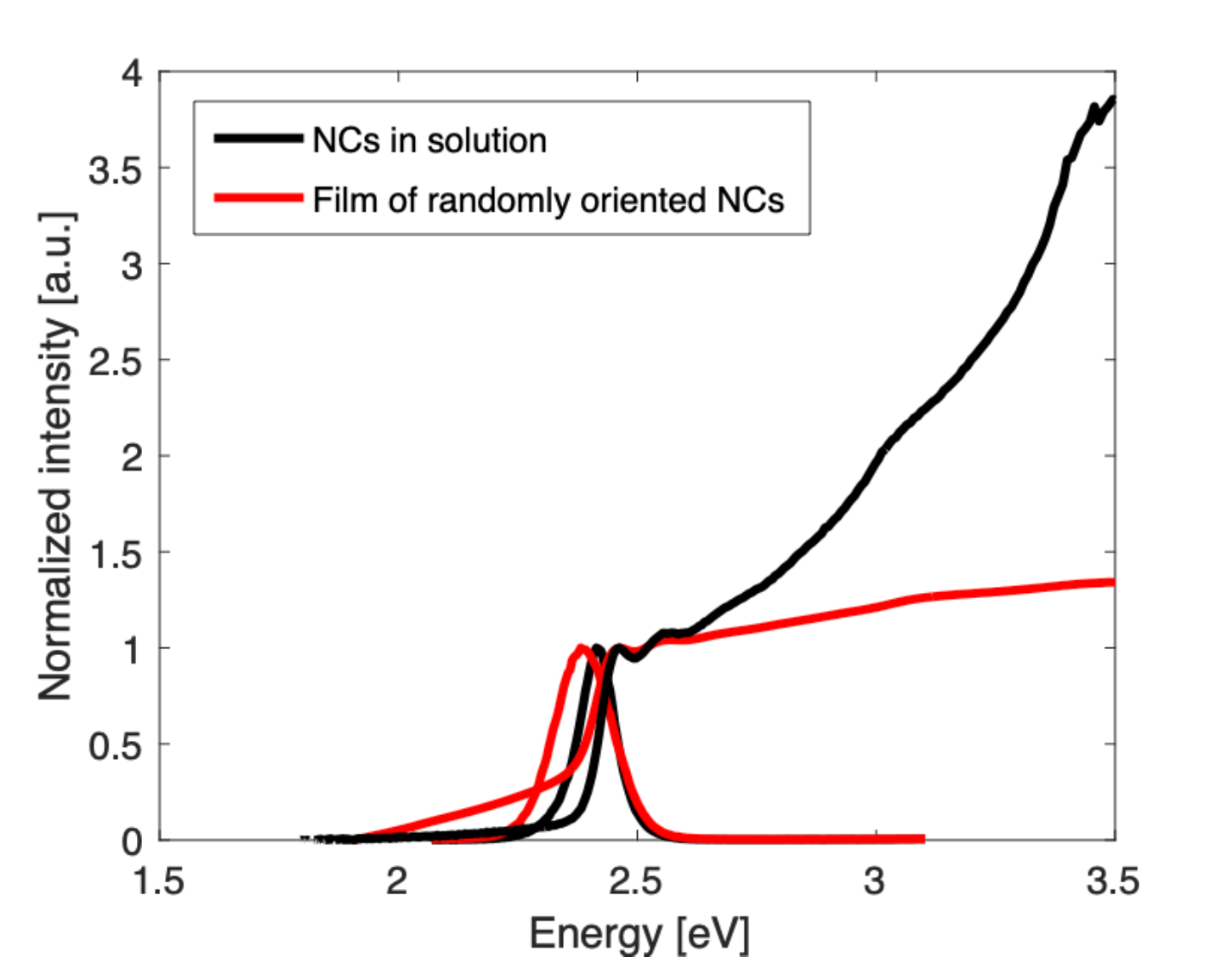}
    \caption[Specimen fixed on the paper graph mask.]{}
\label{SI - fig: PL dis}
    \end{subfigure}
    \begin{subfigure}[t]{0.45\textwidth}
    \includegraphics[width=1\textwidth]{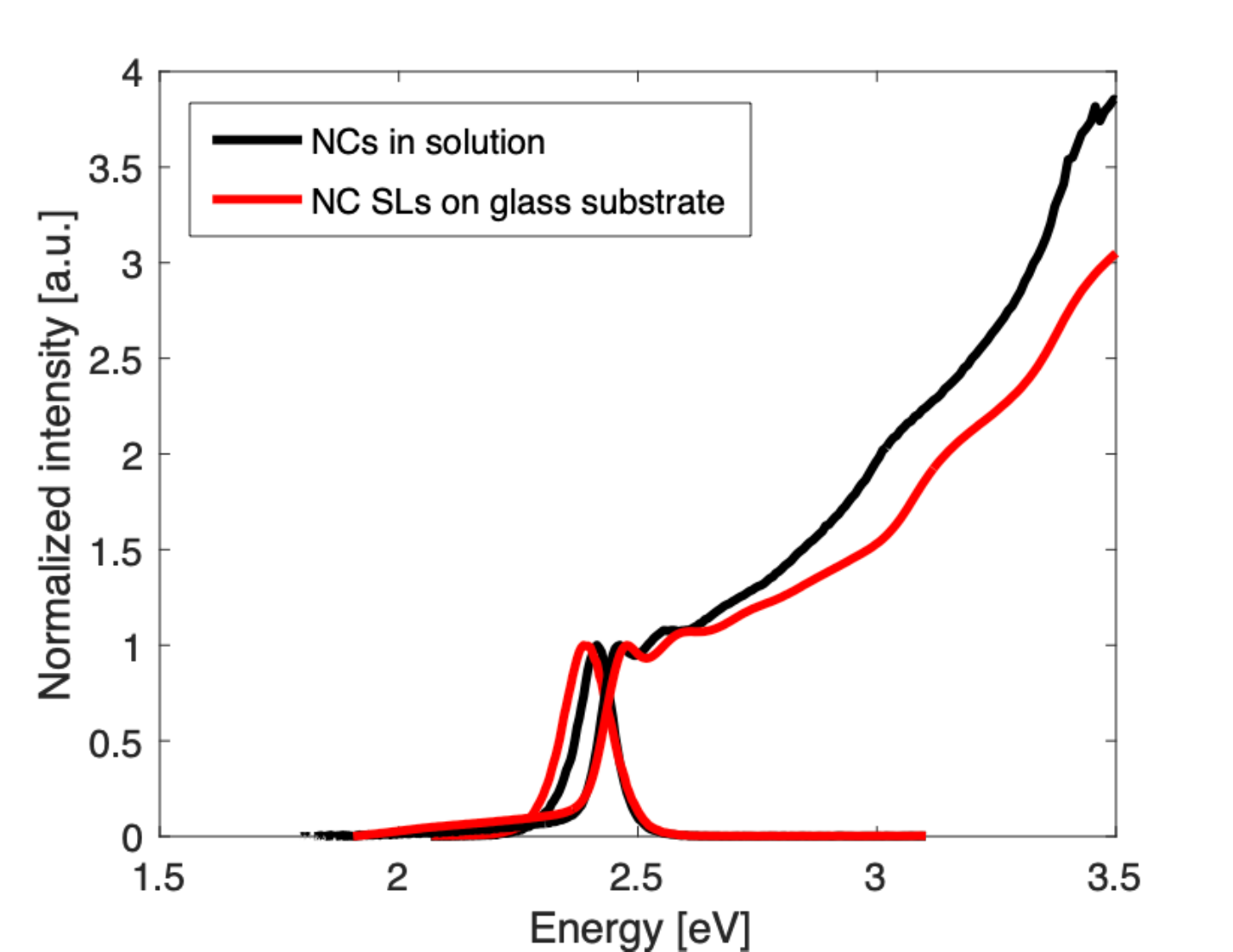}
    \caption[Specimen fixed on the paper graph mask.]{}
    \label{SI - fig: PL SL}
\end{subfigure}
    \caption{(a) Room temperature PL and absorption spectra of NCs dispersed in toluene (black) and of disordered NCs on a glass substrate  (red). (b) Room temperature PL and absorption spectra of NCs dispersed in toluene and of NC SLs deposited on a glass substrate.}
\label{SI - fig: PL}    
\end{figure}

The samples were optically characterized by absorption and photoluminescence (PL) spectra, acquired with a Cary 300 spectrophotometer and with a Cary Eclipse spectrofluorometer,  respectively. The equilibrium PL and absorption spectra of NCs dispersed in toluene and of NC deposited on a glass substrate are displayed in figure \ref{SI - fig: PL dis} and \ref{SI - fig: PL SL} for disordered NC and superlattices respectively. In both cases, we observe a PL red-shift ($\Delta$PL$_{NC} =$ 2.414 eV - 2.382 eV = 32 meV for the disordered NCs sample and $\Delta$PL$_{SL}$ = 2.414 eV - 2.386 eV = 28 meV for the SLs sample) and a PL peak broadening ($\Delta \text{FWHM}_{NC} = 142 ~\text{meV} - 86 ~\text{meV} = 55 ~\text{meV}$ for disordered NCs and $\Delta\text{FWHM}_{SL}$ = 116 meV - 86 meV = 30 meV for SLs) for the deposited films compared to NCs in toluene dispersion, consistently with previous observations \cite{baranov2019investigation,kovalenko2017lead,nagaoka2017nanocube,van2018cuboidal,tong2018spontaneous}.\\

\subsection{X-ray diffraction}
\begin{figure}[h!]
    \centering
     \centering
    \begin{subfigure}[t]{\textwidth}
    \includegraphics[width=1\textwidth]{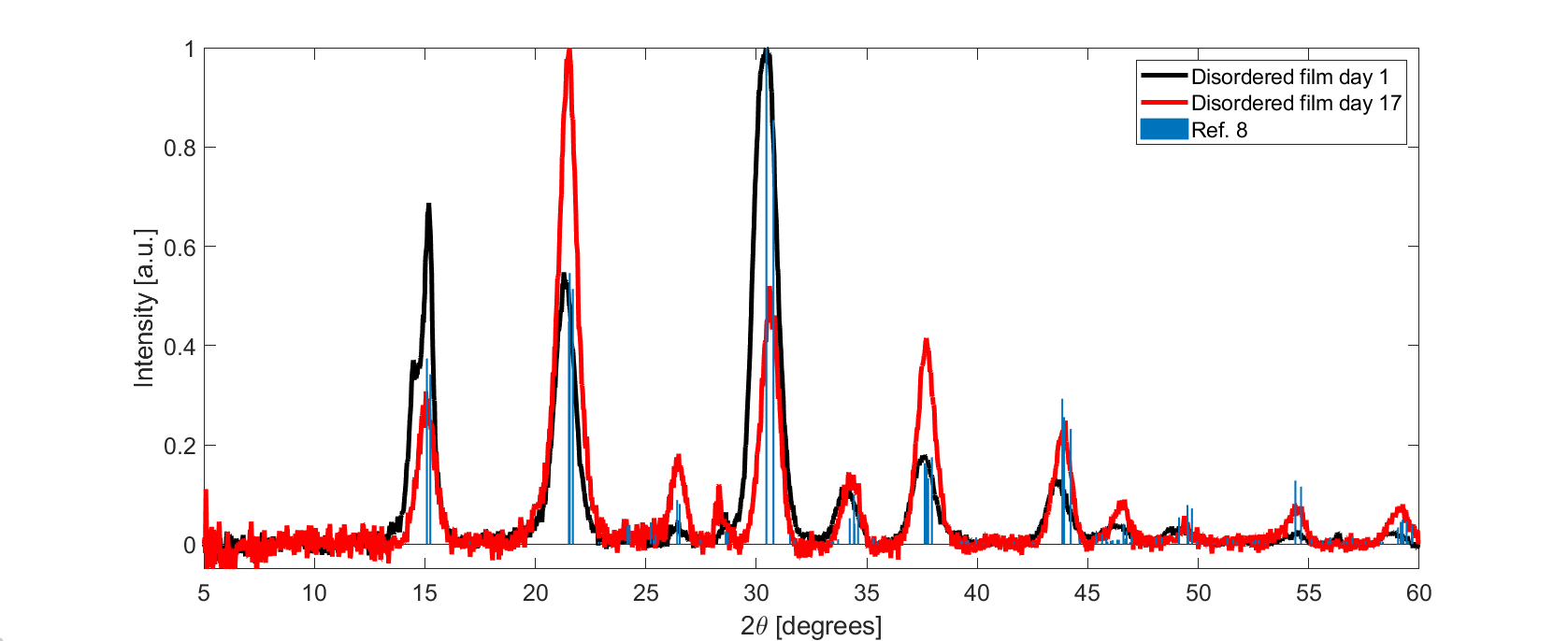} 
    \caption[Specimen fixed on the paper graph mask.]{}
    \label{SI - fig: dis_xrd}
    \end{subfigure}
    \begin{subfigure}[t]{\textwidth}
    \includegraphics[width=1\textwidth]{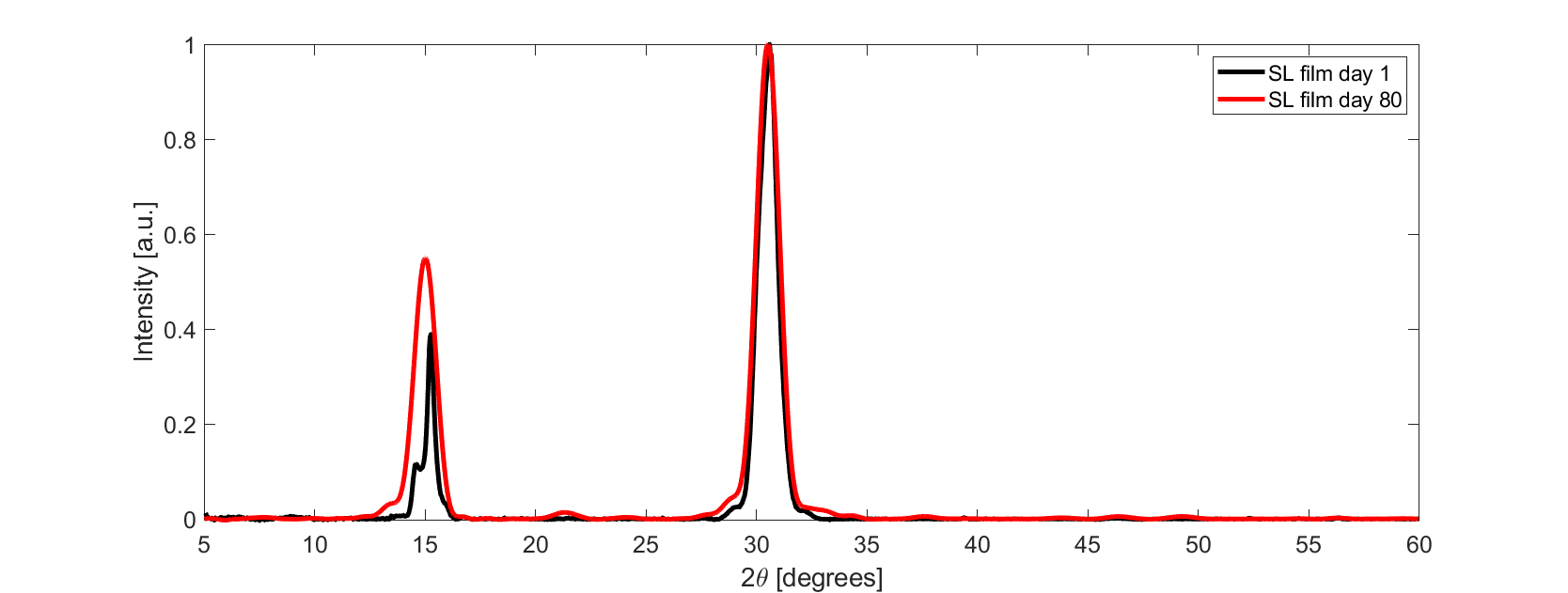}
    \caption[Specimen fixed on the paper graph mask.]{}
    \label{SI - fig: sl_xrd}
    \end{subfigure}
    \caption{XRD patterns of disordered NCs film (a) and superlattice film (b).}
    \label{SI - fig: xld}
\end{figure}

X-ray diffraction (XRD) patterns collected from disordered NCs film and superlattice sample are shown in figure \ref{SI - fig: xld}. The XRD data were measured with a PANalytical Empyrean diffractometer in a parallel beam configuration, equipped with a Cu K$\alpha$ ($\lambda$ = 1.5406 \r{A}) ceramic X-ray tube operating at 45 kV and 40 mA, 1 mm wide incident and receiving slits, and a 40 mA PIXcel3D 2×2 two-dimensional detector.\\
For disordered NCs, the XRD pattern shows the Bragg reflections ascribed to the orthorhombic structure of \ch{CsPbBr_3} \cite{lopez2020crystal} and all the peaks expected from a film of randomly oriented NCs \cite{toso2019wide}. Conversely, for the SL sample, we observe the presence of two strong peaks at $2\theta$ $\sim$ 15$^\circ$ and $2\theta$ $\sim$ 30.5$^\circ$. As described in detail in previous works \cite{toso2019wide,toso2021multilayer}, this XRD pattern originates from a close-packing of the cubes in the plane parallel to the substrate, which leads to the enhanced Bragg reflections from (110), (1$\Bar{1}$0) and (002) ($2\theta$ $\sim$ 15$^\circ$) and (220), (2$\Bar{2}$0) and (004) ($2\theta$ $\sim$ 30.5$^\circ$) planes of the orthorombic unit cell of \ch{CsPbBr_3}. The precise periodicity between NCs inside SLs produces a phase modulation on the diffracted X-rays, causing additional interference which produces the fine structure for the $2\theta$ $\sim$ 15$^\circ$ peak, observable in figure \ref{SI - fig: sl_xrd} (black solid line).

\subsection{Optical microscopy}
\begin{figure}[h!]
    \centering
    \begin{subfigure}[t]{0.45\textwidth}
    \includegraphics[width=1\textwidth]{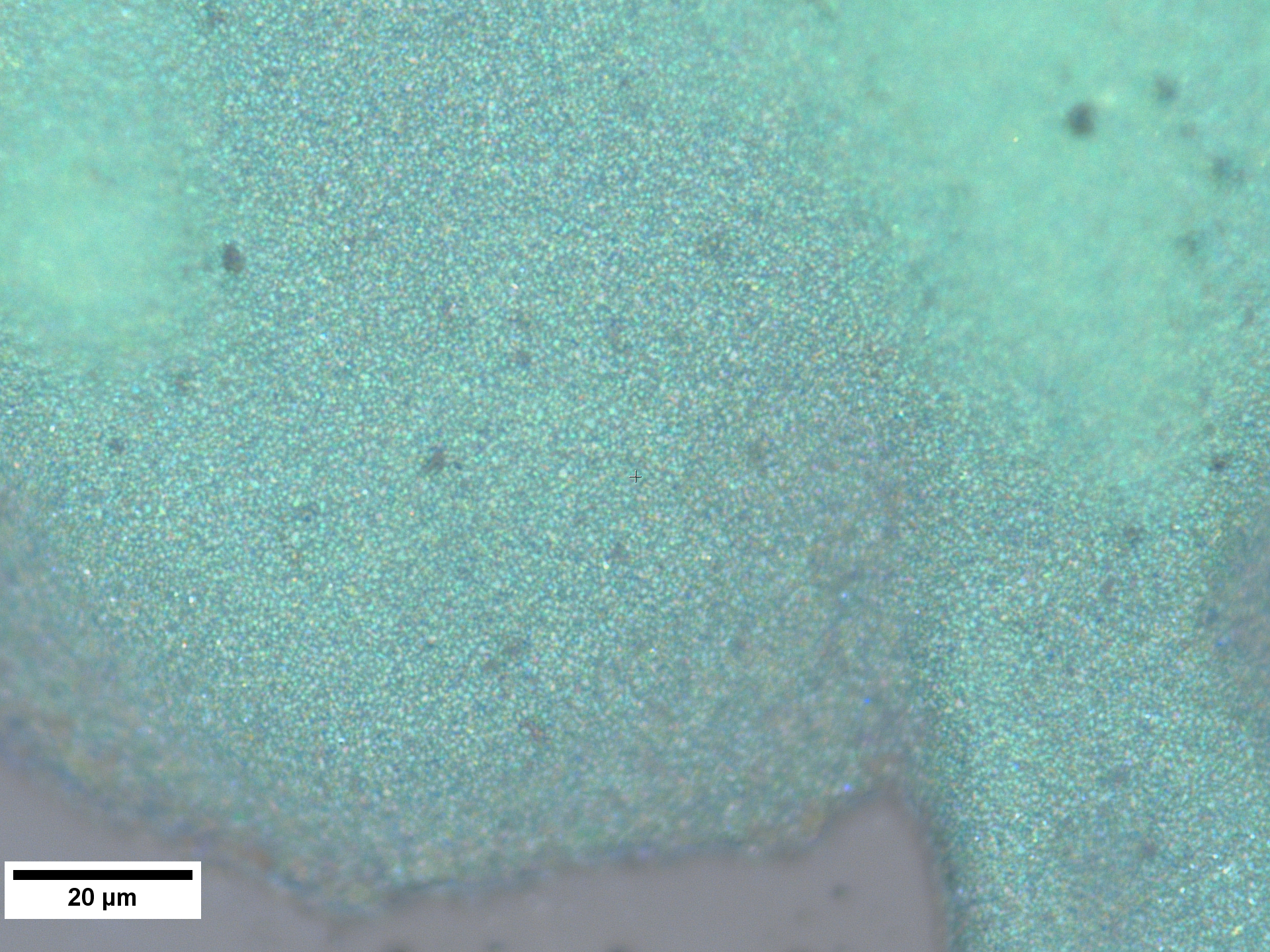}
    \caption[Specimen fixed on the paper graph mask.]{}
    \label{SI - fig: optical microscopy dis}
    \end{subfigure}
    \begin{subfigure}[t]{0.45\textwidth}
    \includegraphics[width=1\textwidth]{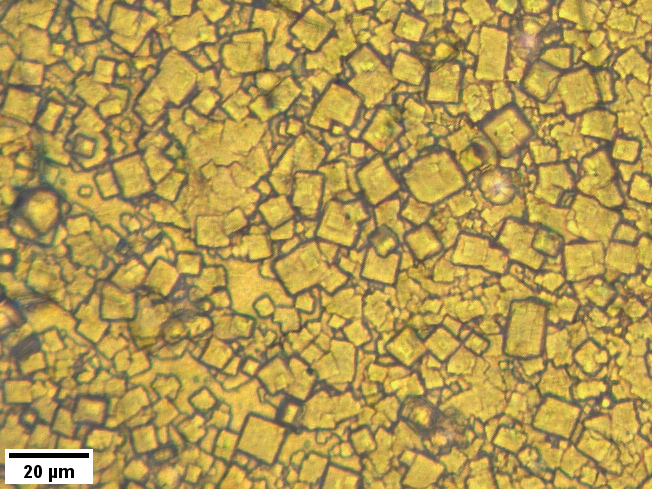} 
    \caption[Specimen fixed on the paper graph mask.]{}
    \label{SI - fig: optical microscopy SL}
    \end{subfigure}
    \caption{Optical microscopy images of disordered NCs film (a) and superlattice film (b).}
    \label{SI - fig: optical microscopy}
\end{figure}

Optical microscope images of the NCs films, acquired on a ZETA-20 true color 3D optical profiler, are displayed in figure \ref{SI - fig: optical microscopy}. For the disordered NCs film (fig. \ref{SI - fig: optical microscopy dis}), the image shows a roughly homogeneous spatial distribution of \ch{CsPbBr_3} NCs; no well-defined micron-sized rectangular structure, that typically corresponds to SLs, is observable in the
sample, suggesting that mechanical scrambling produces largely disordered NC film. In the SL sample (fig. \ref{SI - fig: optical microscopy SL}), the NCs are assembled into aggregates of rectangular shape and size between 1 and 10 \textmu m, each corresponding to a NC superlattice. 

\section{Sample aging}
Pump-probe experiments were performed on samples with various ages, from fresh ($\sim$ day 4) to $\sim$ 90 days old ones. The data presented in the main text were acquired on 30-35 days old disordered NCs sample and 85-92 days old SL sample. When not used for experiments the films were stored in a vacuum desiccator cabinet at $\sim$1 mbar. Aging effects are investigated by means of XRD and optical microscopy.

\subsection{Disordered NCs film}
Figure \ref{SI - fig: dis_xrd} shows the XRD pattern of the fresh (day 1, black solid line) and aged (day 17, red solid line) disordered film. The latter measurement was done after the film had been mounted in the cryostat employed for pump-probe experiments and had been subjected to ambient pressure-to-vacuum cycles and to room temperature-to-17 K cycles. As observable, both XRD patterns show,  with different ratios in the peak intensities, the Bragg reflections ascribed to the orthorhombic structure of \ch{CsPbBr_3}, as expected for a film of disordered NCs. Nevertheless, for the day 1 measurement, the peak at $2\theta$ $\sim$ 15$^\circ$ shows a splitting, which suggests that when the film was deposited, NCs assembled and arranged in a periodic way in some form of super-structures. The size of these superstructures must be $\leqslant$ 1 $\micro$m, since they are not observable with optical microscopy (figure \ref{SI - fig: optical microscopy dis}) and neither with SEM, where just a rough surface is noticeable. After 17 days and after being subjected to several vacuum and temperature cycles, the splitting at the $2\theta$ $\sim$ 15$^\circ$ peak disappears, suggesting that the periodic arrangement is lost. Since all the pump and probe measurements on disordered film were conducted after this last XRD measurement, we can assume that the disordered film we measured is made of randomly oriented and randomly packed NCs.

\subsection{SL film}
\begin{figure}[h!]
    \centering
    \begin{subfigure}[t]{0.48\textwidth}
    \includegraphics[width=1\textwidth]{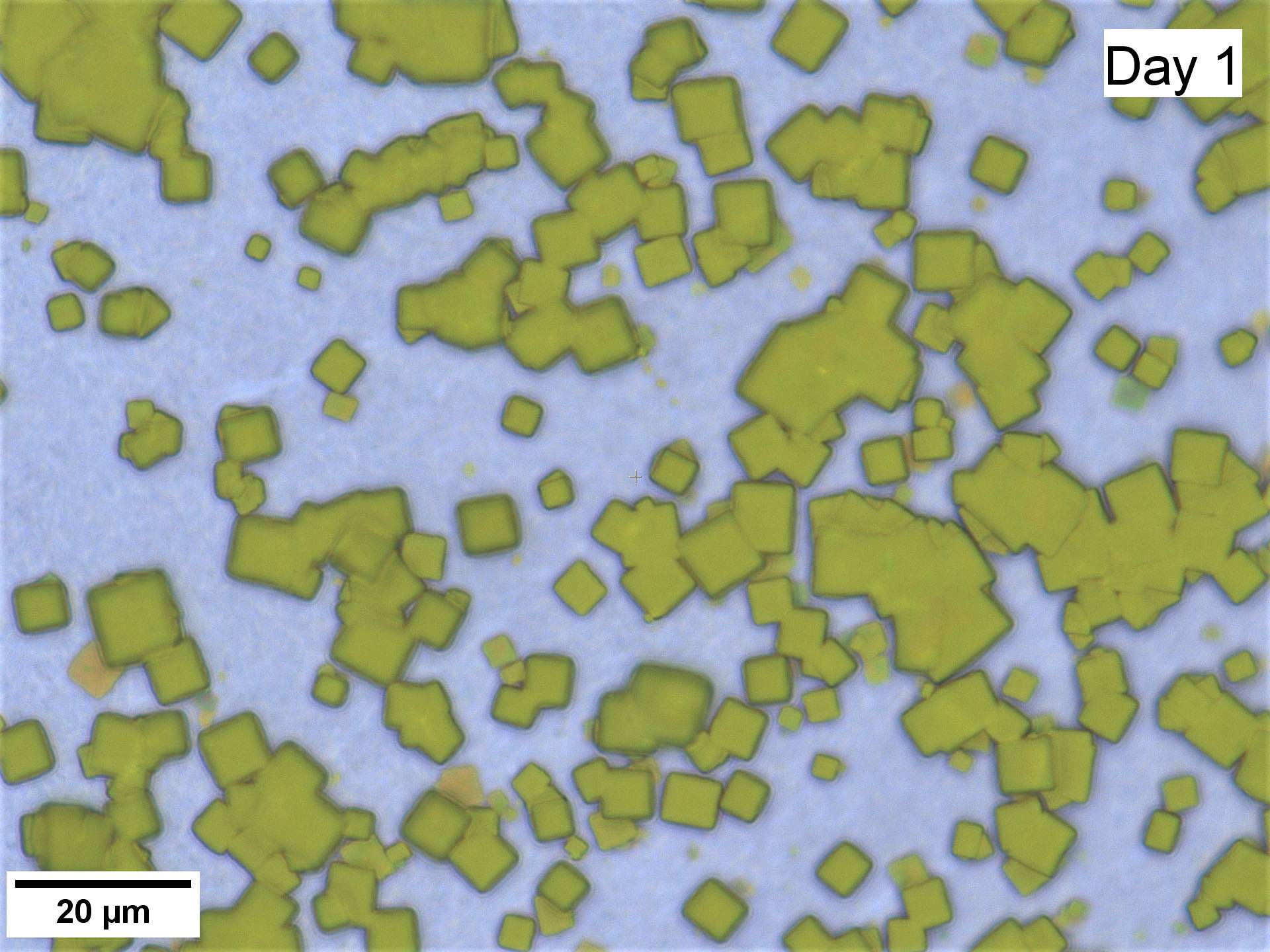}
    \caption[Specimen fixed on the paper graph mask.]{}
    \label{tensile_test_spec}
    \end{subfigure}
    \begin{subfigure}[t]{0.48\textwidth}
    \includegraphics[width=1\textwidth]{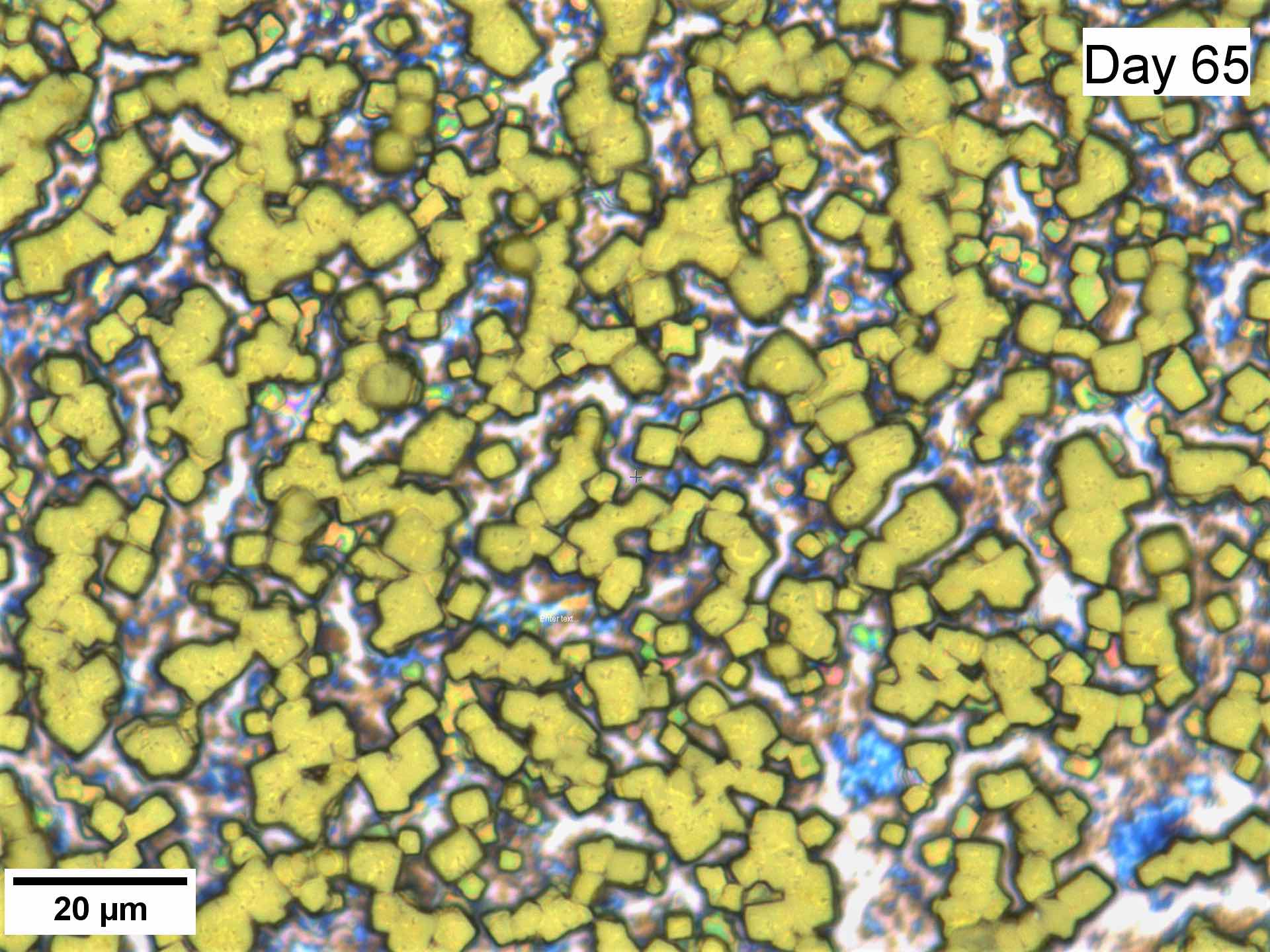}
    \caption[Specimen fixed on the paper graph mask.]{}
    \label{tensile_test_spec}
    \end{subfigure}
    \begin{subfigure}[t]{0.48\textwidth}
    \includegraphics[width=1\textwidth]{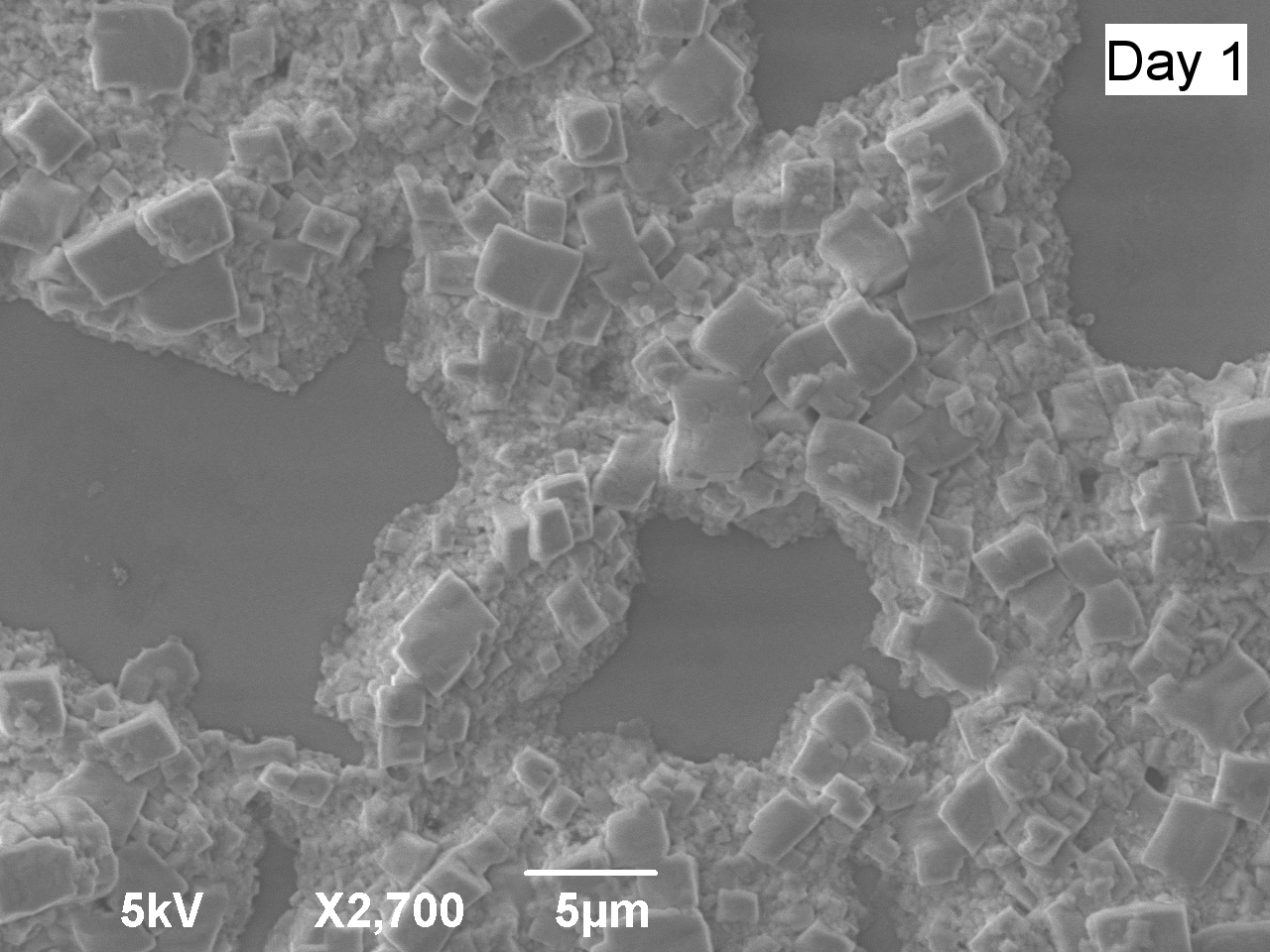}
    \caption[Specimen fixed on the paper graph mask.]{}
\label{tensile_test_spec}
    \end{subfigure}
    \begin{subfigure}[t]{0.48\textwidth}
    \includegraphics[width=1\textwidth]{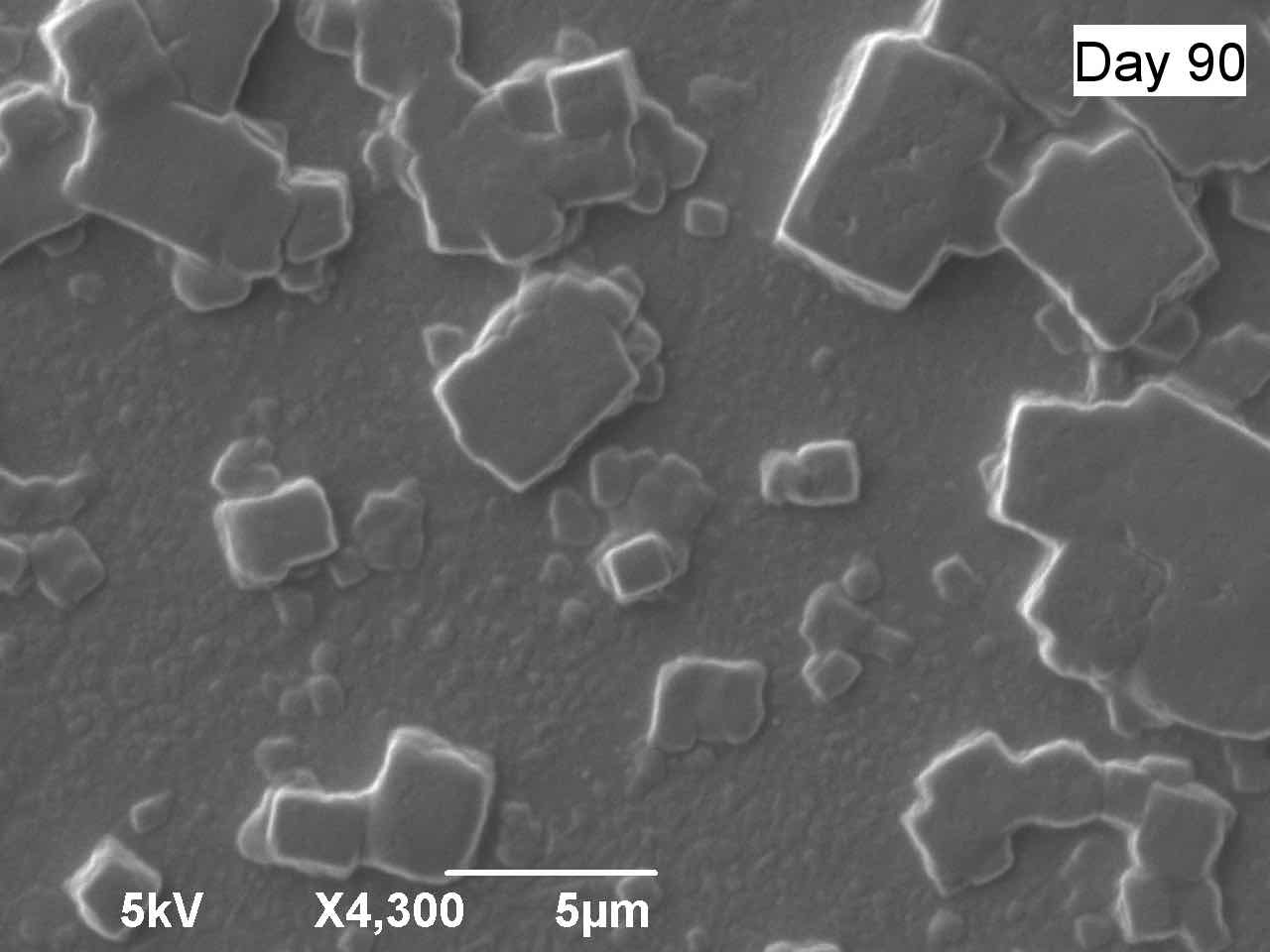}
    \caption[Specimen fixed on the paper graph mask.]{}
    \label{tensile_test_spec}
    \end{subfigure}
    \caption{ Optical microscope and SEM images of fresh (a, c) and aged (b, d) SL films respectively.}
    \label{SI - fig:fresh-aged}
\end{figure}

Figure \ref{SI - fig: sl_xrd} shows XRD patterns of fresh (day 1, black solid line) and aged (day 80, red solid film) films of densely packed \ch{CsPbBr_3} NC SLs. The XRD pattern of the fresh SL film was collected immediately after the evaporation of the solvent ($\approx$ 12 hours after dropcasting solution on Si substrate), while the XRD pattern of the aged SL film was collected after pump and probe experiments, i.e. after the sample had undergone several cycles of vacuum and of cryogenic temperatures. \\
We observe that, also in the aged sample, the main diffraction contribution comes from two peaks ($2\theta$ $\sim$ 15$^\circ$ and $2\theta$ $\sim$ 30$^\circ$), but the fine structure for the $2\theta$ $\sim$ 15$^\circ$ peak is lost upon aging. This superlattice interference is closely related to the statistical fluctuation of the NC spacing, indicated by $\sigma_L$. It represents the stacking disorder of NCs and its effect on the diffraction profile is to smear fringes, making them disappear when its value becomes too high, leaving just one peak profile dependent on atomic plane periodicity and NC thickness. \\
Table \ref{SI - tab: fit} and figure \ref{SI - fig: fit_XRD} summarize the results obtained by fitting the XRD pattern of the fresh and aged films, by means of the multilayer diffraction method described in Ref. \cite{toso2021multilayer}. After 90 days, an increase of $\sigma_L$ from 1.321 \r{A} to 2.322 \r{A} (relative increase of $\sim$ 75$\%$) is observed. This is consistent with the XRD pattern of the aged film shown in figure \ref{SI - fig: sl_xrd}, which displays no fringes at the $2\theta$ $\sim$ 15$^\circ$ peak. 
A decrease of the interparticle spacing L is also observed after 90 days, from 34.2 \r{A} to 29.4 \r{A}. This is attributed to the removal of entrapped solvent (toluene) molecules during the application of vacuum: if we consider the volume of toluene molecule, which is equal to $\sim$177 \r{A}$^3$, and we approximate the volume to the one of a cube, we get an edge equal to $\sim$5.6 \r{A}. That value is comparable to the contraction of L from fresh to aged sample (4.8 \r{A}) obtained from the fit.
It is noteworthy that the average number of atomic planes per NC, N, does not change according to the fit. That implies NCs do not undergo a significant coarsening. \\
Control experiments (not shown) were performed to verify that after applying vacuum the only change in the results of XRD pattern fitting is the decrease of interparticle spacing L due to solvent removal and, subsequently, to verify that upon cooling the sample (to 84 K in this case) no change in L and $\sigma_L$ was observed once the sample was returned to room temperature.\\
These considerations allow us to claim that aging of these samples does not lead to aggregation or side reactions that could form other products \cite{baranov2020aging}. The only aging effect appears to be the increase in disorder manifested as an increase in the inter-NC spacing fluctuation (as depicted in figure \ref{SI - fig: SL_spacing}), due to NC and ligand rearrangement. 
Nevertheless, despite the increased disorder, the aged film is still made of close-packed monodispersed NCs lying in planes parallel to the substrate. The images of the aged film obtained from optical microscopy and scanning electron microscopy (SEM, performed on a JEOL JSM-6490LA microscope) are shown in figure \ref{SI - fig:fresh-aged}b,d and prove that the morphology of aged samples remains similar to fresh ones (fig. \ref{SI - fig:fresh-aged}a,c), with no noticeable impurities on the surface of SLs.\\

\begin{center}
\begin{table}[h]
\caption{\textbf{Fitting parameter of multilayer diffraction model.} Results of XRD pattern fitting  of fresh and aged films shown in figure \ref{SI - fig: fit_XRD}.}
    \centering
\begin{tabular}{c | c | c | c}
\toprule[1.25pt]
    \bf{Parameter} & \bf{Definition} & \bf{Day 1}   & \bf{Day 80} \\ \midrule[1.25pt] \rowcolor[gray]{.95}
d [\r{A}]  & Lattice constant &  5.830       &    5.829    \\ 
\midrule
L [\r{A}]  &  Interparticle spacing  &   34.186      &     29.384                              \\\rowcolor[gray]{.95}
\midrule
$\sigma_L$ [\r{A}]  & Interparticle spacing fluctuation &  1.321     &   2.322                              \\
\midrule
N [atomic planes]  & Number of atomic planes for NC &  14.394   &      14.306                           \\ \rowcolor[gray]{.95}
\midrule
$\sigma_N$ [atomic planes]  & Size distribution of NCs & 1.760       &   2.150                               \\  
\midrule
q-zero correction [\r{A}]  & Correction of diffractometer misalignment&  0.004          &    0.012                               \\ \rowcolor[gray]{.95}
\midrule
NC edge [nm]  & (N-1)$\times$d &  7.8 $\pm$ 0.3 &  7.7 $\pm$ 0.4\\
\bottomrule[1.25pt]
\end{tabular}
\label{SI - tab: fit}
\end{table}
\begin{figure}[h!]
    \centering    
    \begin{subfigure}[t]{\textwidth}
    \includegraphics[width=0.49\textwidth]{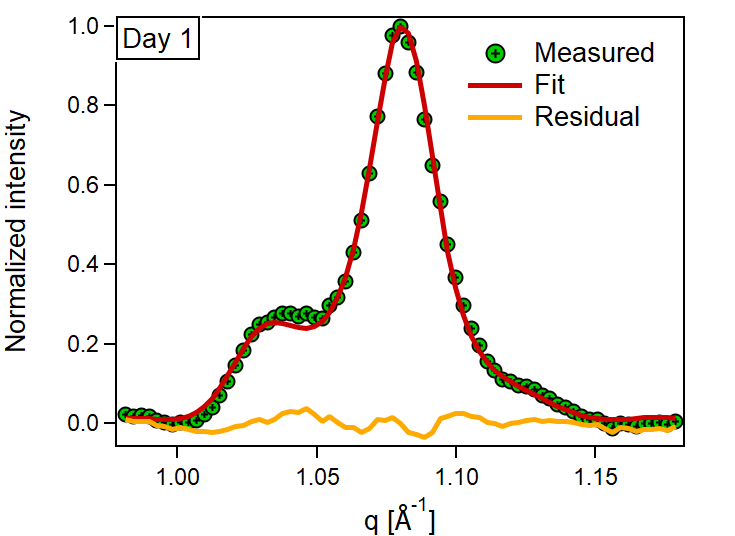}
    \includegraphics[width=0.49\textwidth]{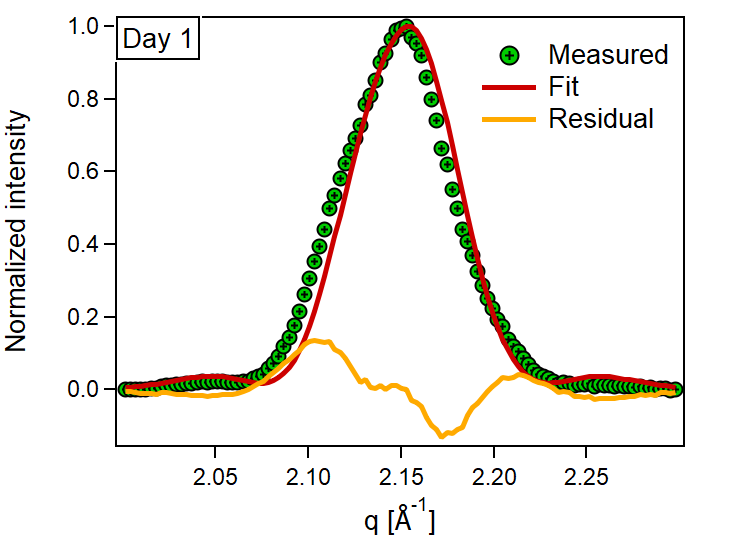}
    \caption[Specimen fixed on the paper graph mask.]{}
    \label{SI - fig: fit_fresh}
    \end{subfigure}
    \begin{subfigure}[t]{\textwidth}
    \includegraphics[width=0.49\textwidth]{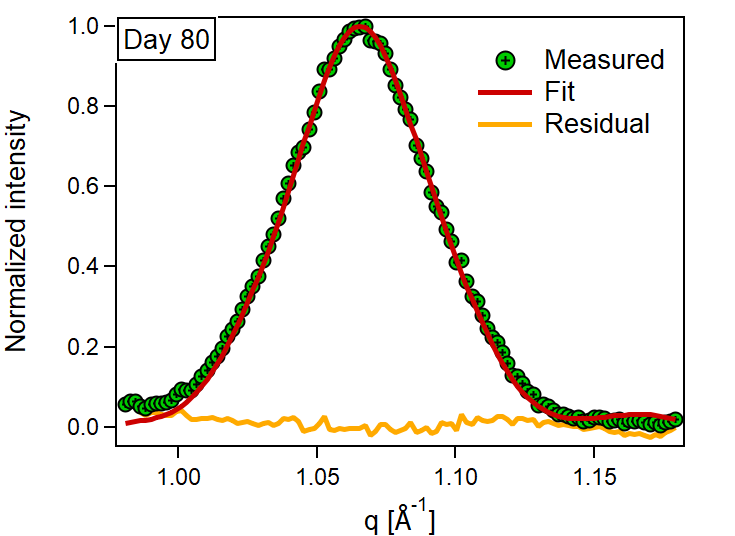}
    \includegraphics[width=0.49\textwidth]{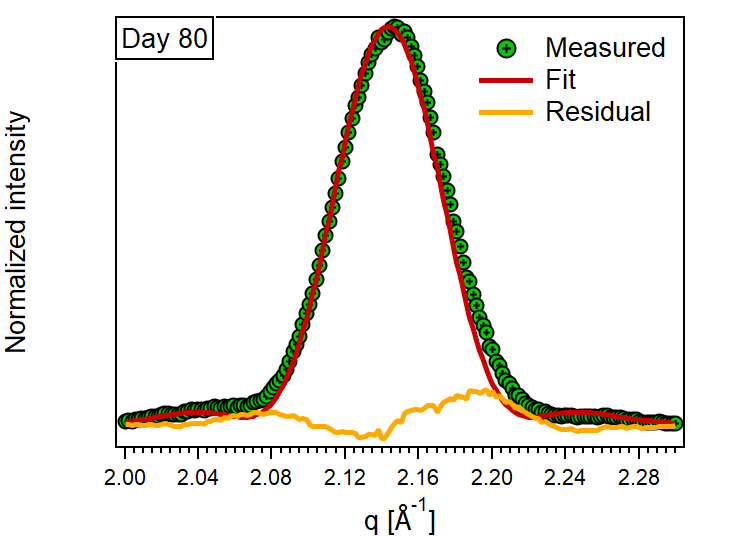}
    \caption[Specimen fixed on the paper graph mask.]{}
    \label{SI - fig: fit_aged}
    \end{subfigure}
    \caption{Fitting of $2\theta$ $\sim$ 15$^\circ$ and $2\theta$ $\sim$ 30.5$^\circ$ (q $\sim$ 1.065, 2.111 \r{A}$^{-1}$ respectively, q$\,\, = (4\pi/\lambda_{exc})$sin$(\theta)$) XRD peaks of fresh (a) and aged (b) films performed by means of the multilayer diffraction model described in Ref. \cite{toso2021multilayer}.}
    \label{SI - fig: fit_XRD}
\end{figure}
\end{center}

\begin{figure}[h!]
     \centering
    \includegraphics[width=0.9\textwidth]{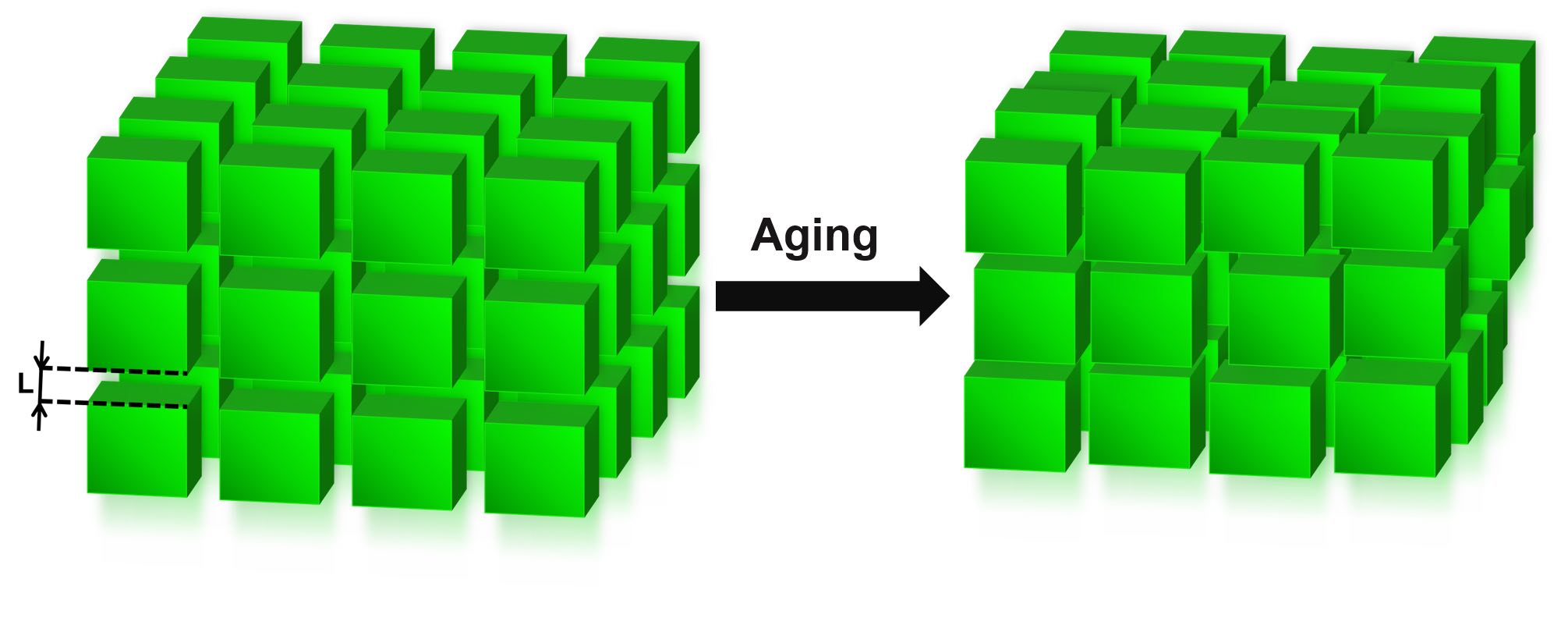}
    \caption{Cartoon of the main effect of aging that causes a decrease of NC spacing in SLs and an increase of NC spacing fluctuations. The NC close packing arrangement and preferential orientation is nevertheless preserved.}
    \label{SI - fig: SL_spacing}
\end{figure}

\clearpage
\section{Equilibrium optical properties}
\label{SI - sec: equilibrium optical properties}

Analysis of the pump-probe data by means of differential fitting of $\Delta R/R$ spectra requires a parametrization of the \ch{CsPbBr_3} optical constants at low temperature (T). We therefore build the real and imaginary parts of the low-T refractive index by employing a suitable Kramers-Kr\"onig consistent model, where the relevant parameters are chosen in agreement with the energy values obtained from the experimental absorbance measured at room temperature (fig. \ref{SI - fig: PL SL}) and with the temperature dependent trends reported in literature. 

\begin{figure}[h!]
\centering
\includegraphics[width=16cm]{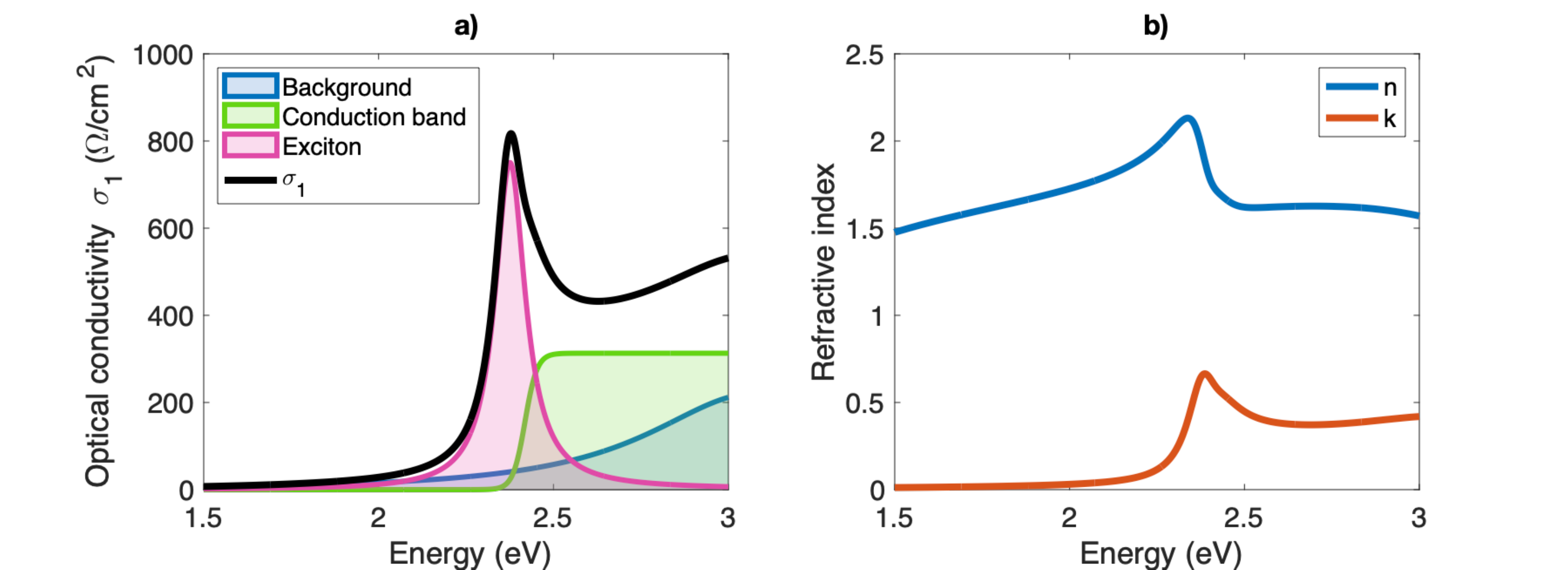}
\caption{a) Parametrized real part of the optical conductivity at low temperature (black line). Filled areas represent the different components included in $\sigma_1$. b) Real 
($n$) and imaginary ($k$) parts of the parameterized low temperature refractive index.}
\label{SI - fig: equilibrium optical properties}
\end{figure}

The real part of the optical conductivity, $\sigma_1$, is obtained as the sum of a sigmoid function, accounting for the conduction band edge, a Drude-Lorentz oscillator, describing the exciton resonance, and a background component modelled through a high-energy Drude-Lorentz oscillator (figure \ref{SI - fig: equilibrium optical properties}a). The conduction band gap energy is red-shifted by $\sim$80 meV as compared to room temperature (fig. \ref{SI - fig: PL SL}) and the amplitude of the exciton peak is increased compared to the edge amplitude, consistently with the temperature dependence of \ch{CsPbBr_3} absorption spectra reported in literature \cite{guo2019dynamic,shcherbakov2021temperature}. The exciton binding is fixed at 43 meV, in agreement with literature reports \cite{protesescu2015nanocrystals,li2016temperature} and experimental absorbance (fig. \ref{SI - fig: PL SL}). The imaginary part $\sigma_2$ is then computed from Kramers-Kro\"nig relations. The resulting real and imaginary parts of the refractive index ($n$ and $k$, respectively) are plotted in figure \ref{SI - fig: equilibrium optical properties}b. We note that the results of the differential fitting procedure reported in the main text and in Sec. \ref{S_pumpprobe} are robust if small changes of the equilibrium dielectric function are introduced. 

\section{Fit of pump-probe data}
\label{S_pumpprobe}
\begin{figure}[h!]
\centering
\includegraphics[width=14.5cm]{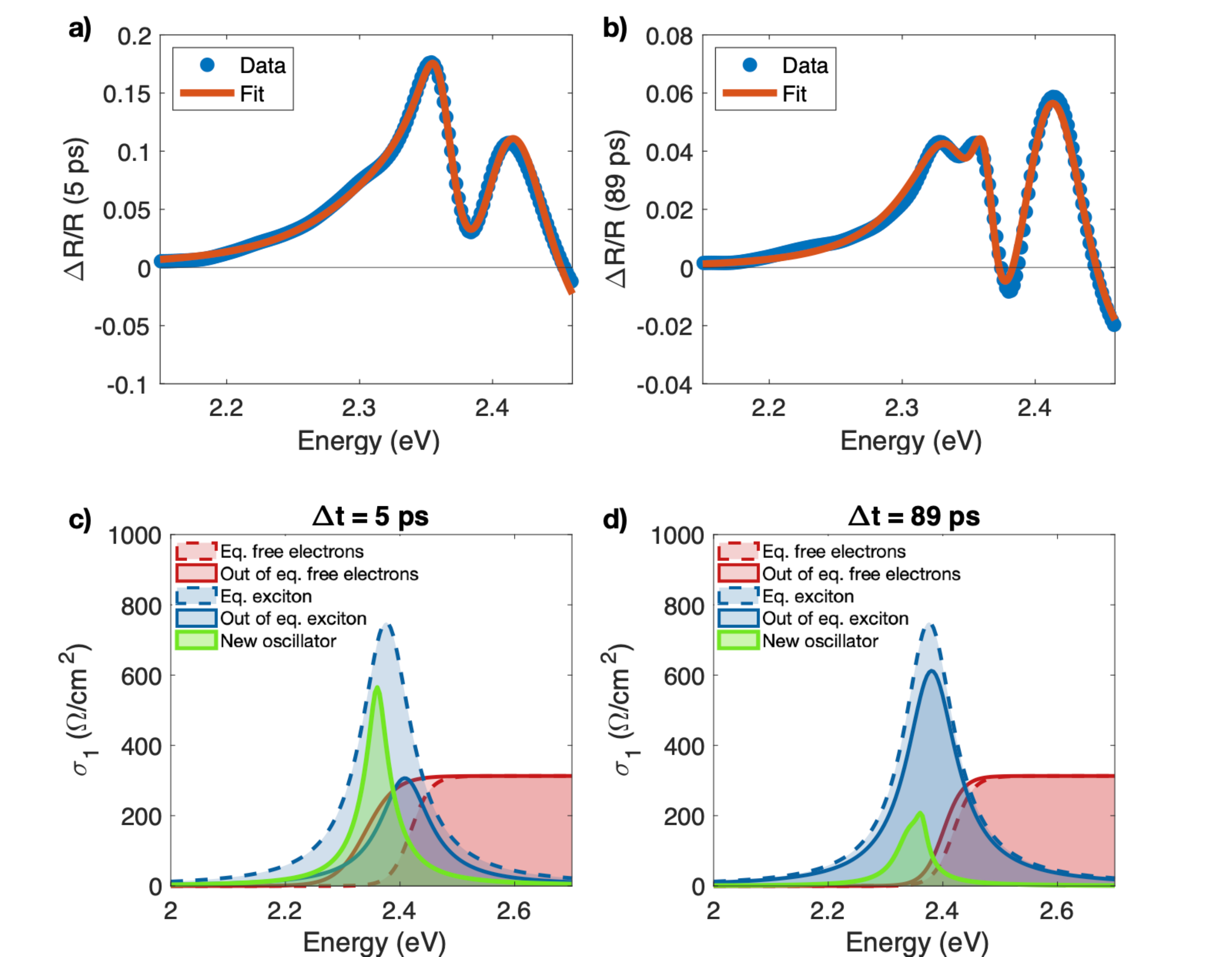}
\caption{Differential fit of $\Delta R/R$ data measured on NCs superlattices at 17 K with 230 \textmu J/cm$^2$ excitation fluence. a) and b) show the fitted spectra at 5 ps and 89 ps respectively. c) and d) report the equilibrium (dashed lines) and out of equilibrium (solid lines) components of the optical conductivity.}
\label{SI - fig: fit high F}
\end{figure}

For each measured pump-probe time delay $\Delta t$ between 4 ps and 130 ps, the spectrally resolved reflectivity variation $\Delta R/R$ is fitted with a function
\begin{equation}
    \frac{\Delta R}{R} = \frac{R_{outeq}-R_{eq}}{R_{eq}}
\end{equation}
where $R_{eq}$ is the equilibrium reflectivity estimated as $((1-n)^2+k^2)/((1+n)^2+k^2)$, $n$ and $k$ being the real and imaginary parts of the refractive index plotted in figure \ref{SI - fig: equilibrium optical properties}b. $R_{eq}$ is evaluated as the normal incidence bulk reflectivity because the light penetration depth in the photon energy range of interest is $\lesssim$ 1 \textmu m, which is smaller than the film overall thickness (on the order of $\sim$ 10 \textmu m).  $R_{outeq}$ is the out-of-equilibrium reflectivity, computed with the same procedure employed for $R_{eq}$ and containing the fit parameters. The free parameters for the differential fit are the out-of-equilibrium free-carriers edge energy position, the sigmoid edge width, the exciton energy and the exciton spectral weight, which represents the smallest subset of free parameters necessary to reproduce the out of equilibrium reflectivity at all fluences and all timescales. The data collected from NC superlattices require to include in the out-of-equilibrium optical conductivity also an additional feature at energy $\sim$2.36 eV, which is smaller then the exciton resonance. Since, at long time delays ($\Delta t \gtrsim$ 50 ps), the pump-probe map in figure 2b shows two distinct peaks between 2.30 eV and 2.37 eV, the experimental data is best described by modelling the new feature below band gap as a sum of two Drude-Lorentz oscillators centered at slightly different energies. This double-peak behaviour likely originates from local inhomogeneities of the samples within the photo-excited area, where superlattices of different lateral size or assembled from NCs of different size can be present and contribute to the measured signal. Figure \ref{SI - fig: fit high F} a and b show the the best differential fit at short (5 ps) and long (89 ps) time delays respectively, as reference examples. Fig. \ref{SI - fig: fit high F} c and d report the various components of the corresponding out-of-equilibrium optical conductivity (solid lines) obtained from the best fit to the data. 

\begin{figure}[h!]
\centering
\includegraphics[width=16cm]{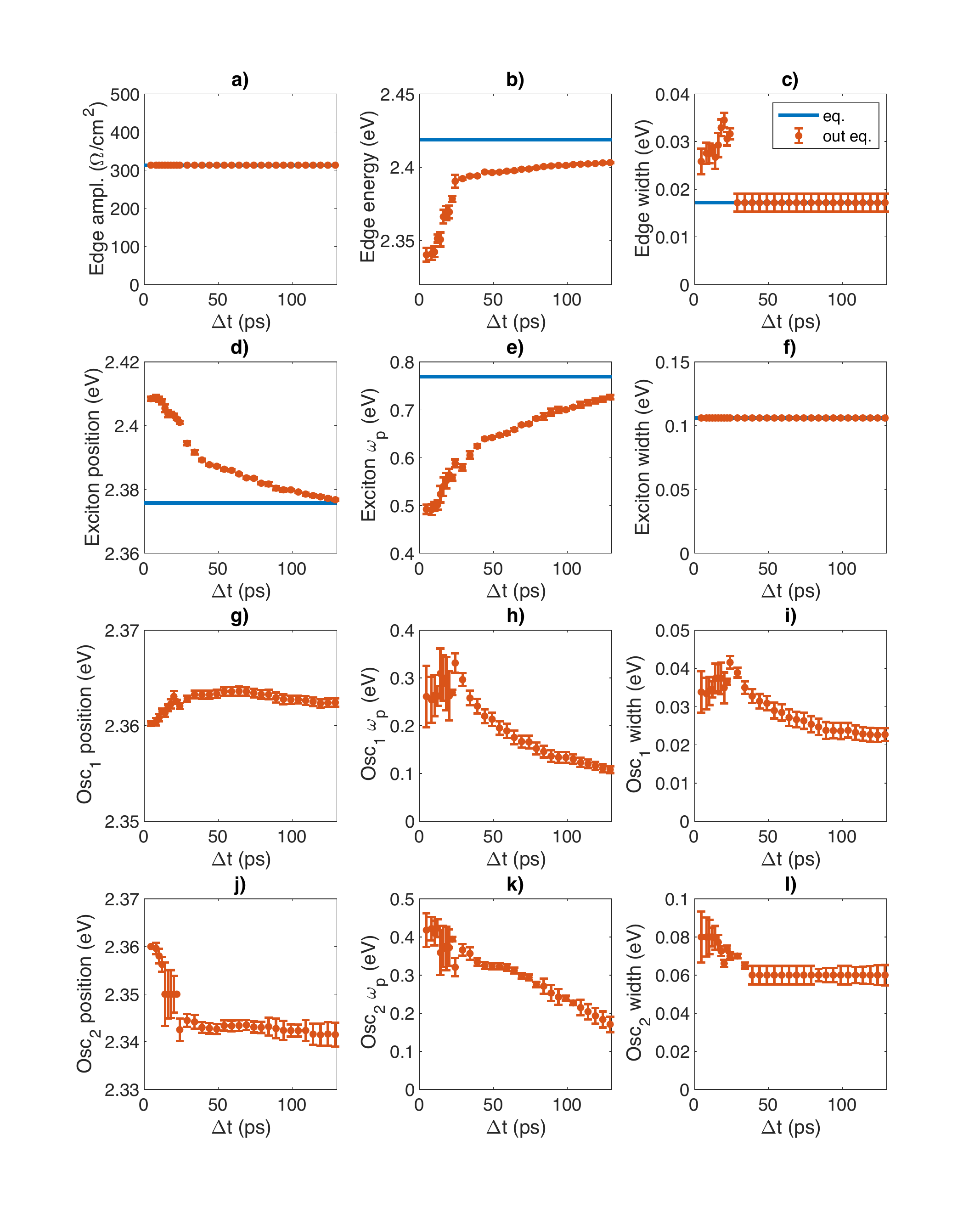}
\caption{$\Delta R/R$  fit output parameters for the superlattice NCs sample at 230 \textmu J/cm$^2$ excitation fluence. a), b) and c) are the free-carriers edge amplitude, center energy and width, respectively. d), e) and f) represent the exciton parameters: exciton energy, plasma frequency and width, respectively. g)-l) display the Drude-Lorentz model parameters for the new oscillators.}
\label{SI - fig: fit high F dynamics}
\end{figure}

\newpage
The output parameters of the fit performed at all time delays are displayed in figure \ref{SI - fig: fit high F dynamics}, where the blue solid lines represent the values at equilibrium and the red markers are the out-of-equilibrium values extracted from the fit. The top row reports the free-carriers sigmoid edge parameters: the edge amplitude (fig. \ref{SI - fig: fit high F dynamics}a)) is kept constants at all $\Delta t$; the red-shift (fig. \ref{SI - fig: fit high F dynamics}b)) observed after pump excitation decays with two distinct dynamics, a fast one ($\sim$ 20 ps) and a lower one taking place over hundreds of ps timescale; the edge width displays a broadening in the first $\sim$ 30 ps. The second row in fig. \ref{SI - fig: fit high F dynamics} shows the dynamics of the parameters of the Drude-Lorentz oscillator relative to the excitonic line, namely the exciton energy (fig. \ref{SI - fig: fit high F dynamics}d)), the plasma frequency $\omega_p$ (fig. \ref{SI - fig: fit high F dynamics}e)) and the exciton width (fig. \ref{SI - fig: fit high F dynamics}f). The exciton width does not show significant changes during the relaxation dynamics and is therefore kept constant at all $\Delta t$ to increase the stability of the fitting procedure. The two bottom rows in figure \ref{SI - fig: fit high F dynamics} report the parameters relative to the Drude-Lorentz oscillators appearing out of equilibrium and associated to the emergence of the cooperative behaviour discussed in the main text. Whereas one single new oscillator is needed to qualitatively reproduce the spectral feature around 2.36 eV at short time delays, two distinct oscillators allow to fit the more structured response observed at long time delays. For consistency, we employed the same model with two new oscillators also at short time delays, where the effect of two distinguished components is covered by the free-carriers edge shift and therefore the error-bars of the associated parameters are larger. 

It is  relevant to point out that the details of the model employed to fit the $\Delta R/R$ have only a marginal effect on the overall result of the fitting procedure. In addition to what presented in figure \ref{SI - fig: fit high F dynamics}, we tested out other models that involved either only one new oscillator, no free-carriers edge width variation, a narrower edge width at equilibrium, or different starting points for the parameters in figures \ref{SI - fig: fit high F dynamics}g)-l). Whatever the fit function employed, the extracted parameters and their relative dynamics display always compatible values and similar trends: the red-shift of the free-carriers edge decaying with two distinct dynamics, the blue-shift and the quench of the exciton recovering to the equilibrium values over $\sim$100 ps and the appearance of new oscillators that are red-shifted and narrower than the equilibrium excitonic peak. 

\begin{figure}[h!]
\centering
\includegraphics[width=8cm]{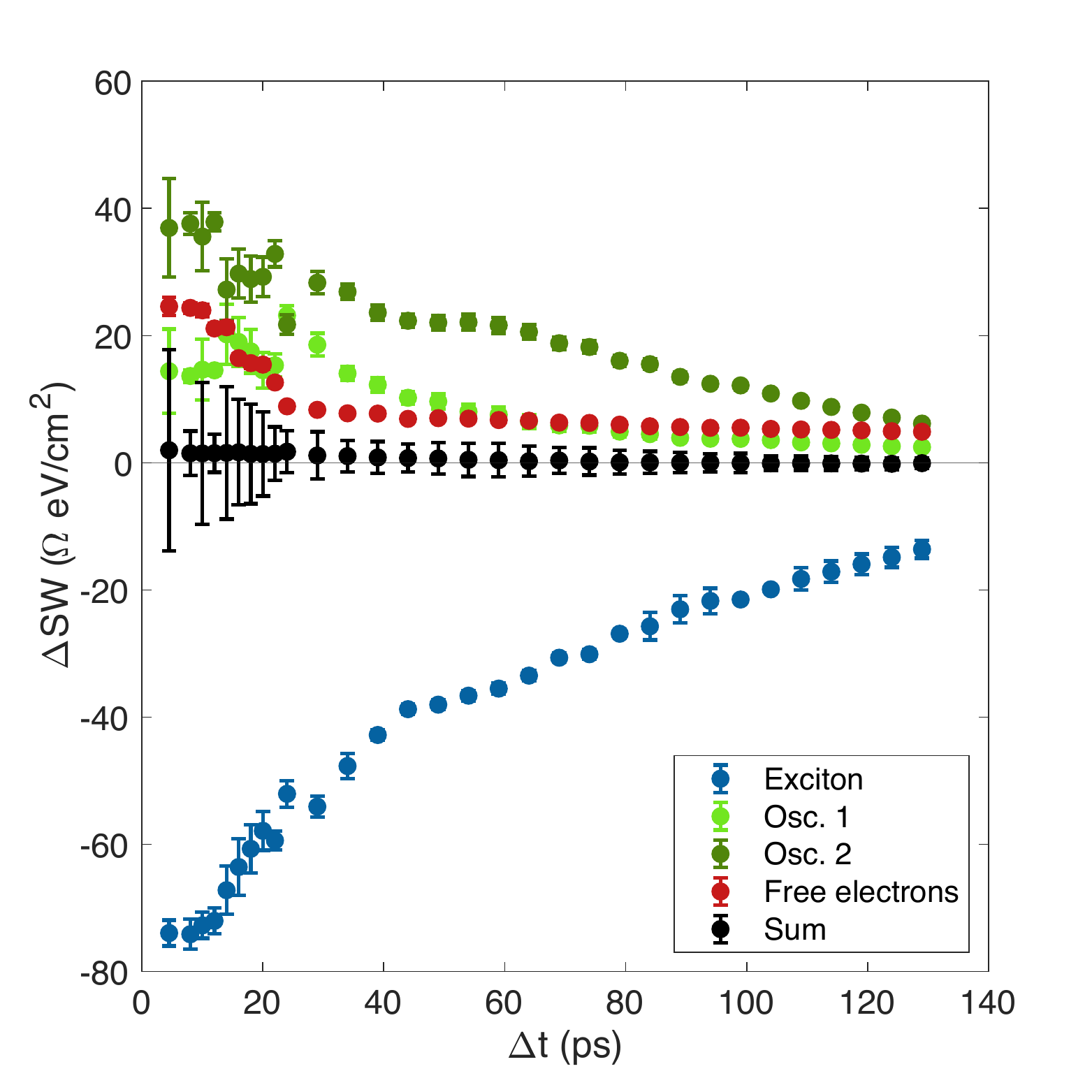}
\caption{Spectral weight variation of the exciton (blue), new oscillators (green), and conduction band (red) after pump excitation. Black points represent the sum of the four contributions, indicating SW conservation.}
\label{SI - fig: SW}
\end{figure}

Inspection of the spectral weight transfer between the exciton resonance and the free-carrier band is performed by computing the spectral weight variation $\Delta$SW of the different components coming into play. 
Figure \ref{SI - fig: SW} reports $\Delta$SW of each component, estimated as the integral of the optical conductivity $\sigma_1$ over the energy axis. We observe that the SW lost after photo-excitation by the exciton peak is re-distributed between the new oscillators and free-carrier states in conduction band. Overall, the spectral weight is conserved, as shown by the black markers in figure \ref{SI - fig: SW} representing the sum of all SW variation contributions, which is compatible with zero at all time delays.

\begin{figure}[h!]
\centering
\includegraphics[width=8cm]{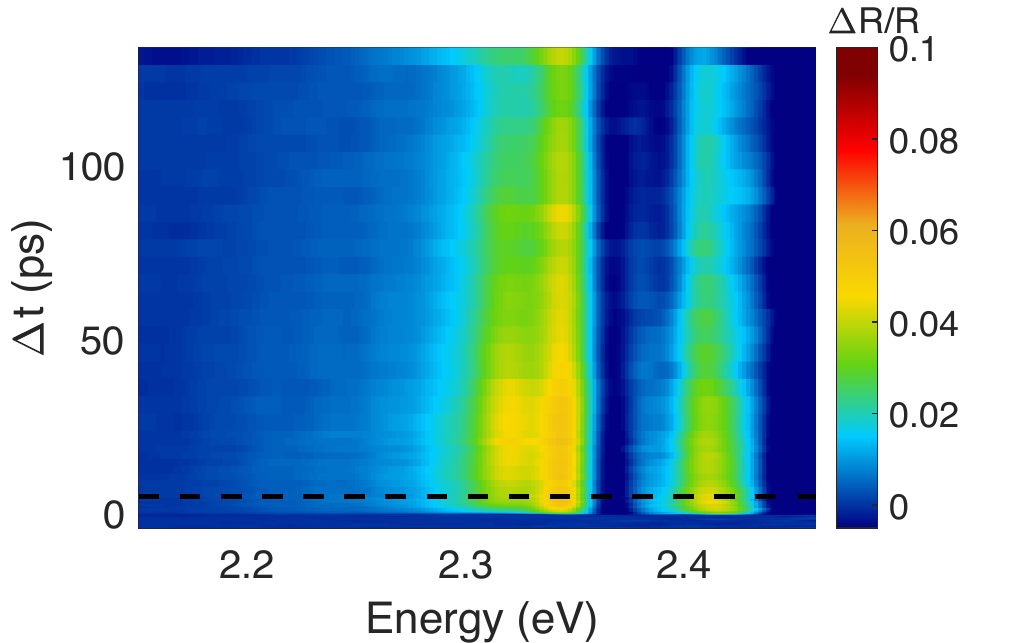}
\caption{Ultrafast transient reflectivity of \ch{CsPbBr_3} superlattice sample measured at 17 K and low excitation fluence (30 \textmu J/cm$^2$).}
\label{SI - fig: low F}
\end{figure}

\begin{figure}[h!]
\centering
\includegraphics[width=14.5cm]{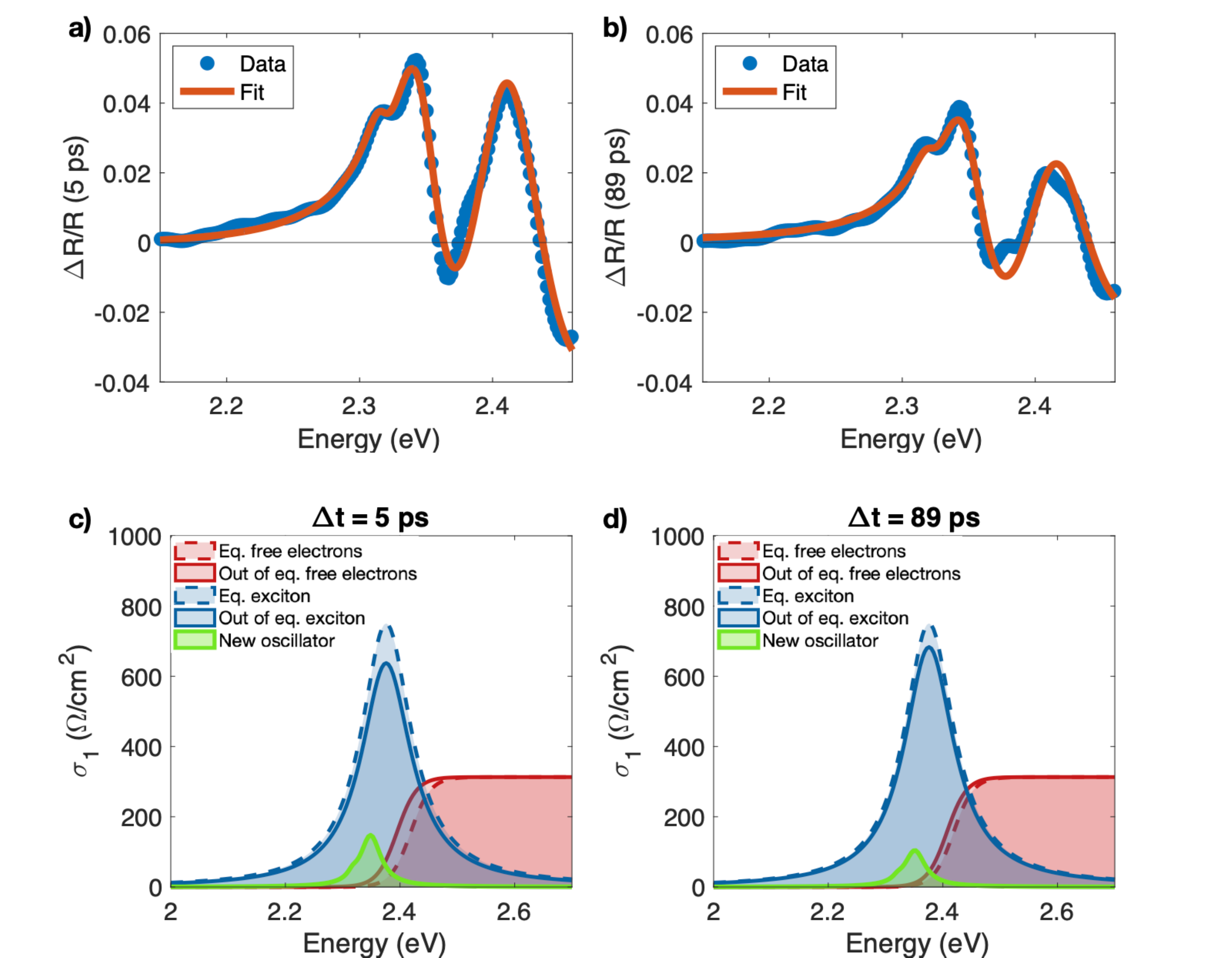}
\caption{Differential fit of $\Delta R/R$ data measured on NCs superlattices at 17 K with 30 \textmu J/cm$^2$ excitation fluence. a) and b) show the fitted spectra at 5 ps and 89 ps respectively. c) and d) report the equilibrium (dashed lines) and out of equilibrium (solid lines) components of the optical conductivity.}
\label{SI - fig: fit low F}
\end{figure}

\begin{figure}[h!]
\centering
\includegraphics[width=16cm]{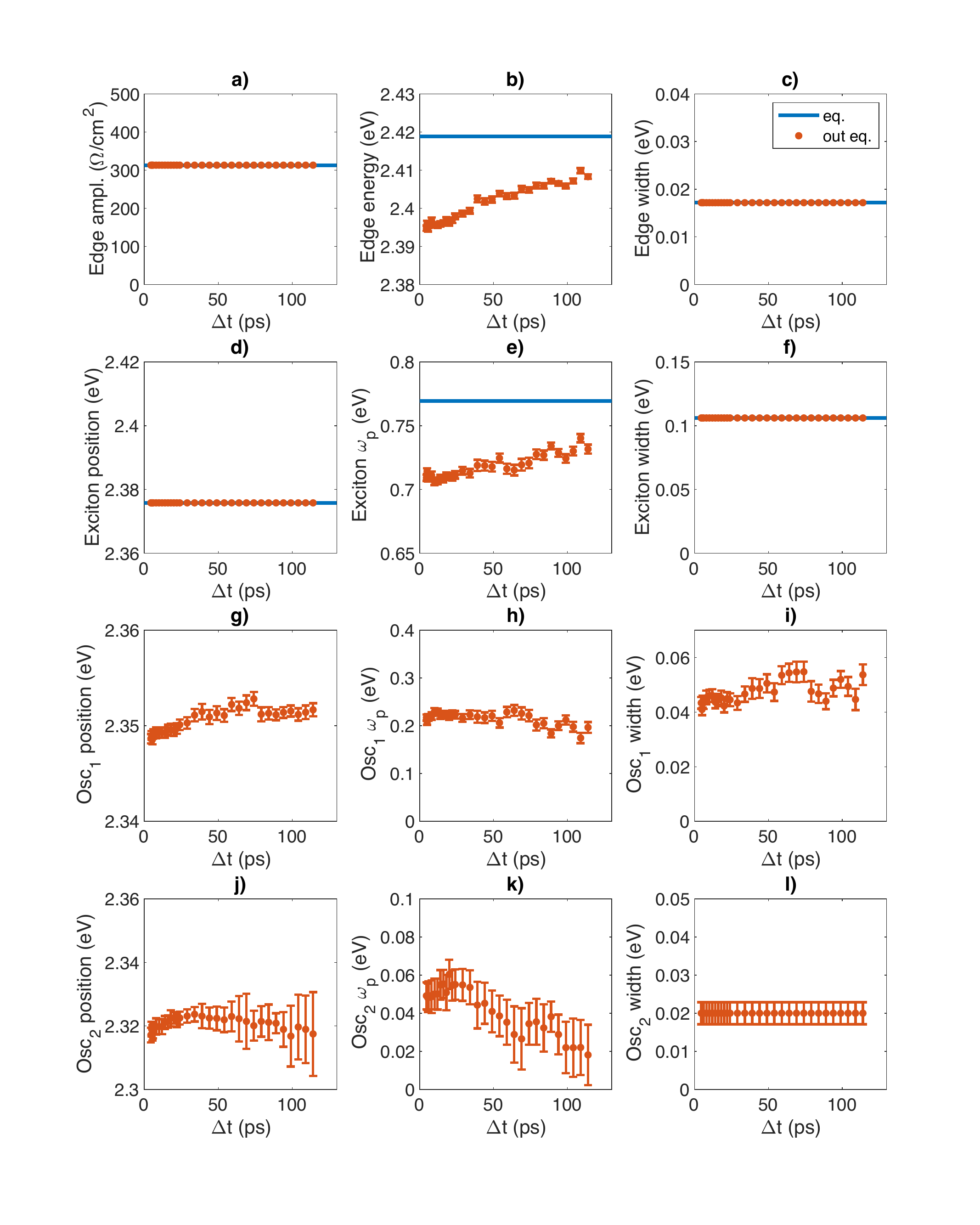}
\caption{$\Delta R/R$  fit output parameters for the superlattice NCs sample at 30 \textmu J/cm$^2$ excitation fluence. a), b) and c) are the free-carriers edge amplitude, center energy and width, respectively. d), e) and f) represent the exciton parameters: exciton energy, plasma frequency and width, respectively. g)-l) display the Drude-Lorentz model parameters for the new oscillators.}
\label{SI - fig: fit low F dynamics}
\end{figure}

\newpage
The same fitting procedure is applied to the data collected on the same superlattice sample with lower excitation fluence (30 \textmu J/cm$^2$), plotted in figure \ref{SI - fig: low F}. The fit results are displayed in figure \ref{SI - fig: fit low F} for two time delays (5 ps and 89 ps) and in figure \ref{SI - fig: fit low F dynamics} (fit output parameters for all time delays). In this lower fluence excitation scheme the free-electron states edge red-shift is smaller and relaxes with a slow dynamics of 130 ps decay time. The exciton does not show any energy shift, but its spectral weight decreases and is transferred to new oscillators appearing out of equilibrium. Since the free-electrons edge amplitude and width as well as the exciton position and width do not show any significant variation during the relaxation dynamics, they are kept fixed to increase the fit stability.

\begin{figure}[h!]
\centering
\includegraphics[width=14.5cm]{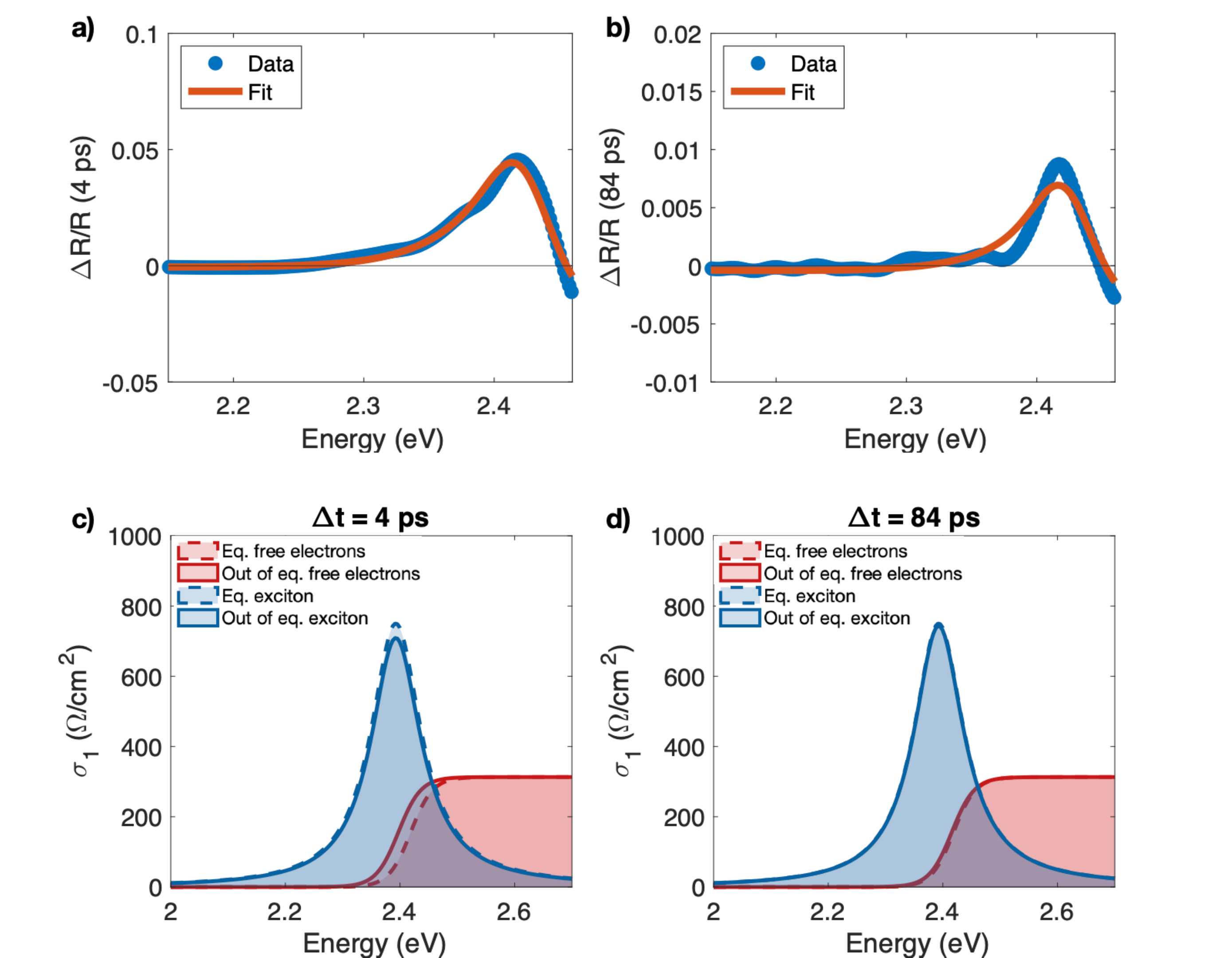}
\caption{Differential fit of $\Delta R/R$ data measured on disordered NCs at 17 K. a) and b) show the fitted spectra at 4 ps and 84 ps respectively. c) and d) report the equilibrium (dashed lines) and out of equilibrium (solid lines) components of the optical conductivity.}
\label{SI - fig: fit dis}
\end{figure}

\begin{figure}[h!]
\centering
\includegraphics[width=16cm]{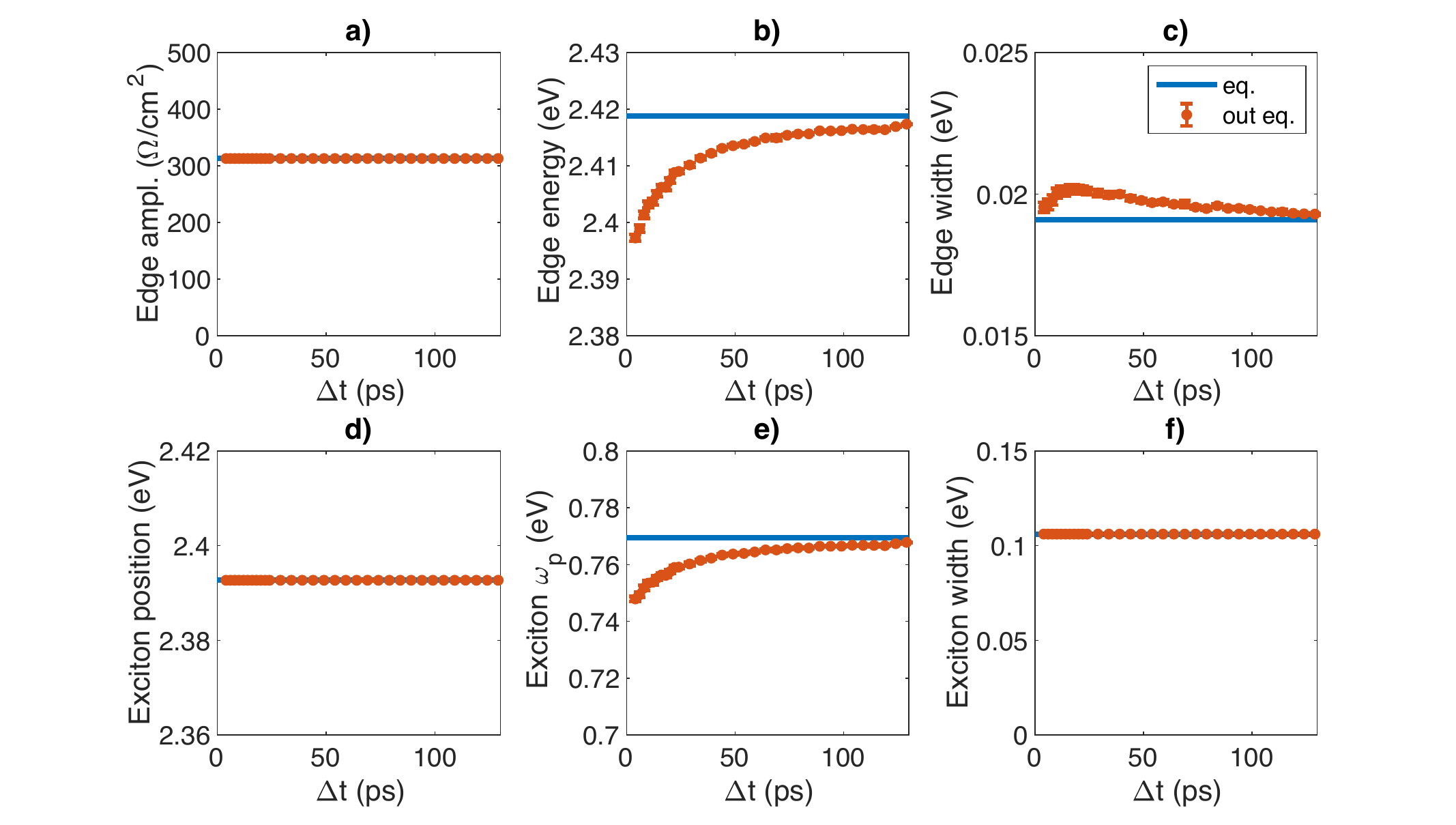}
\caption{$\Delta R/R$  fit output parameters for the disordered NCs sample. a), b) and c) are the free-carriers edge amplitude, center energy and width, respectively. d), e) and f) represent the exciton parameters: exciton energy, plasma frequency and width, respectively.}
\label{SI - fig: fit dis dynamics}
\end{figure}

\newpage
For the disordered NCs sample ($\Delta R/R$ data in fig. 2, the fit output is reported in figures \ref{SI - fig: fit dis}  and \ref{SI - fig: fit dis dynamics}. 
In this case, no additional oscillator is needed to reproduce the out of equilibrium optical conductivity. After pump excitation, a red-shift of the free-carriers edge decaying over a $\sim$ 20 ps timescale is revealed, along with a broadening of the edge width and a decrease of the exciton spectral weight. No significant variation of the free-electrons edge amplitude and of the exciton position and width is observed. 

\clearpage
\section{Fluorescence}
\begin{figure}[h!]
\centering
\includegraphics[width=8cm]{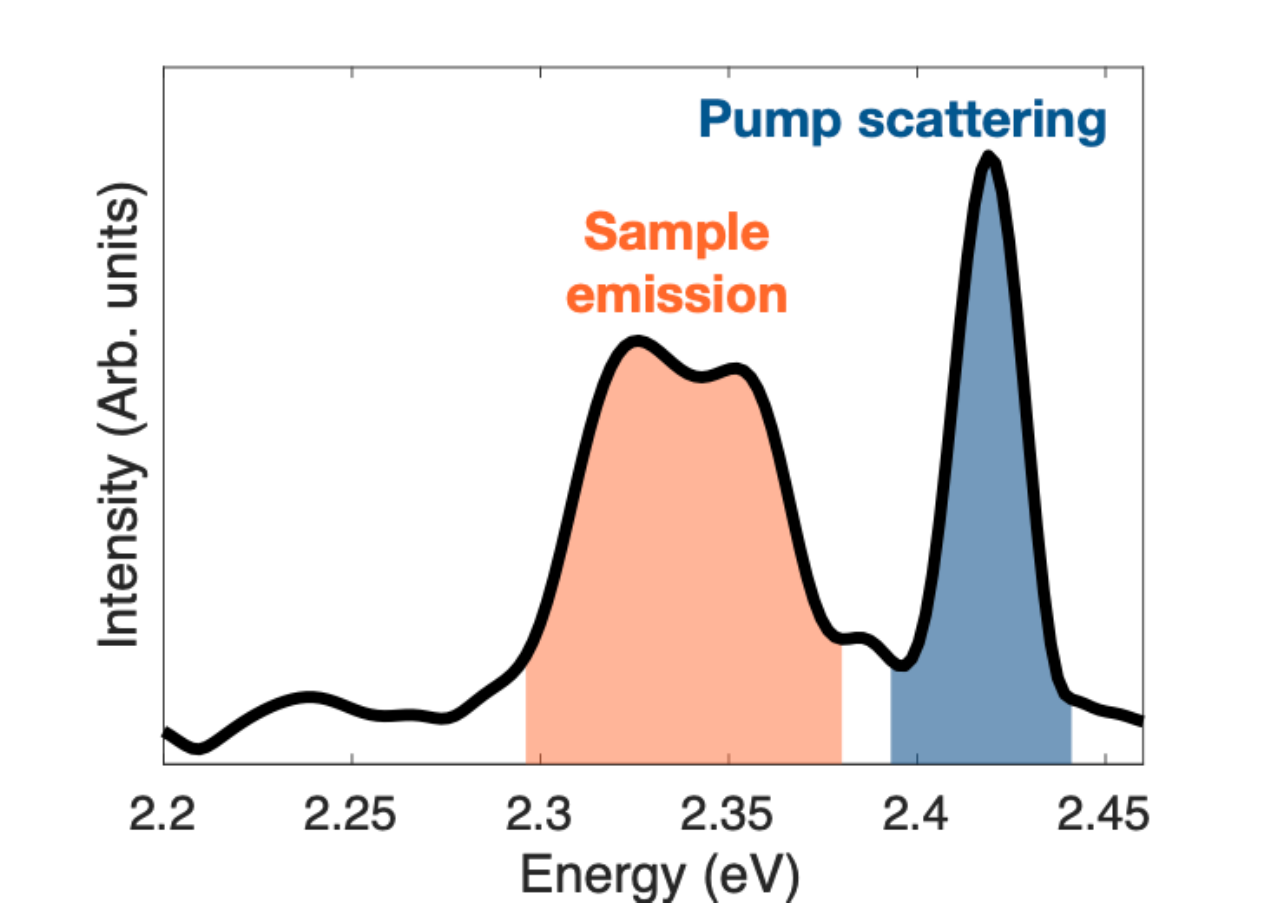}
\caption{Spectrum of the stray light detected when the sample is illuminated by the pump beam only.}
\label{SI - fig: fluorescence}
\end{figure}

Figure \ref{SI - fig: fluorescence} displays the  spectrum of the stray light coming at the detector when the pump pulse only impinges onto the sample. The narrow high energy component is associated to the scattering of the pump beam (blue region in Fig. S9). The radiation detected in the spectral range between 2.30 eV and 2.38 eV (orange region in Fig. S9) comes from the \ch{CsPbBr_3} superlattice sample and is associated to fluorescence emission.

\section{Exciton density}
\label{section: mean N}
In order to estimate the number of excitons generated by the pump pulse in each individual nanocube, we start from the equilibrium optical properties of \ch{CsPbBr_3} (figure \ref{SI - fig: equilibrium optical properties}) at the pump photon energy and calculate the reflectivity $R$, which results being $\approx12\%$. Since the penetration depth is much smaller than the sample thickess, we assume that all the radiation that is not reflected is absorbed. The number of absorbed photons per unit area is then given by
\begin{equation}
    N_{\gamma/area} = \frac{F}{\hbar \omega} (1-R)
\end{equation}
where $\hbar \omega$ is the photon energy and $F$ is the incident fluence. The incident photons are absorbed within a thickness of 135 nm, which corresponds to the penetration depth $l_p$ at the pump photon energy. 
The number of absorbed photons per unit volume is then estimated from
\begin{equation}
    N_{\gamma/volume} = \frac{N_{\gamma/area}}{l_p}.
\end{equation}
Lastly, the number of absorbed photons per nanocube can be obtained by taking into account the volume of each NC, modelled as a cube of $L $ = 8 nm side: 

\begin{equation}
    N_{\gamma/NC} = N_{\gamma/volume} L^3.
\end{equation}

We now assume that each absorbed photon produces an electronic excitation, of which $75\%$ are electron-hole bound states and $25\%$ are free carriers. These values are estimated from the contributions of exciton and band edge to the equilibrium optical conductivity at 17 K. Given the values of pump fluence employed in our experiment, we estimate to generate a number of excitons in each NC that can range between $0$ and $25$. Taking into account that the number of perovskite unit cells within each NC is $L^3/a^3$, $a$ = 5.83 \r{A}, this excitation regime corresponds to a photodoping level up to $\sim1\%$.


\bibliography{Refs}